\documentclass[pdftex,citeautoscript,aps,pra,superscriptaddress,cite,amsmath,amsfonts,amssymb,twocolumn]{revtex4}
\pdfoutput=1
\renewcommand{\vec}[1]{\mathbf{#1}} 
\usepackage{graphicx}
\usepackage{color}
\usepackage{url}

\newcommand{\figref}[1]{Fig.~\ref{fig:#1}}

\newcommand{\Figref}[1]{Figure~\ref{fig:#1}}

\newcommand{\eqnumref}[1]{(\ref{eq:#1})}
\renewcommand{\eqref}[1]{Eq.~\eqnumref{#1}}

\newcommand{\citeasnoun}[1]{Ref.~\onlinecite{#1}}
\newcommand{\citeasnouns}[1]{Refs.~\onlinecite{#1}}
\newcommand{\secref}[1]{Sec.~\ref{sec:#1}}

\newcommand{\Appref}[1]{Appendix~\ref{sec:#1}}

\begin{document}
\title{Structural anisotropy and orientation-induced Casimir repulsion in fluids}
\author{Alexander~P.~McCauley}
\affiliation{Department of Physics, Massachusetts Institute of Technology, Cambridge MA 02139, USA}
\author{F.~S.~S.~Rosa}
\affiliation{Laboratoire Charles Fabry, Institut d'Optique, CNRS, Universit\'{e} Paris-Sud, Campus Polytechnique, RD128, 91127 Palaiseau Cedex, France}
\affiliation{Theoretical Division, Los Alamos National Laboratory, Los Alamos, New Mexico 87545, USA}
\author{Alejandro~W.~Rodriguez}
\affiliation{Department of Mathematics, Massachusetts Institute of Technology, Cambridge MA 02139, USA}
\affiliation{School of Engineering and Applied Sciences, Harvard University, Cambridge MA 02139, USA}
\author{John~D.~Joannopoulos}
\affiliation{Department of Physics, Massachusetts Institute of Technology, Cambridge MA 02139, USA}
\author{D.~A.~R.~Dalvit}
\affiliation{Theoretical Division, Los Alamos National Laboratory, Los Alamos, New Mexico 87545, USA}
\author{Steven~G.~Johnson}
\affiliation{Department of Mathematics, Massachusetts Institute of Technology, Cambridge MA 02139, USA}

\date{\today}

\begin{abstract}
In this work we theoretically consider the Casimir force between two
periodic arrays of nanowires (both in vacuum, and on a substrate
separated by a fluid) at separations comparable to the period.
Specifically, we compute the dependence of the exact Casimir force
between the arrays under both lateral translations and rotations.
Although typically the force between such structures is
well-characterized by the Proximity Force Approximation (PFA), we find
that in the present case the microstructure modulates the force in a
way qualitatively inconsistent with PFA.  We find instead that
effective-medium theory, in which the slabs are treated as
homogeneous, anisotropic dielectrics, gives a surprisingly accurate
picture of the force, down to separations of half the period.  This
includes a situation for identical, fluid-separated slabs in which the
exact force changes sign with the orientation of the wire arrays,
whereas PFA predicts attraction.  We discuss the possibility of
detecting these effects in experiments, concluding that this effect is
strong enough to make detection possible in the near future.
\end{abstract}
\maketitle

\section{Introduction}

Casimir forces are usually attractive interactions measurable at small
separations, but recent theoretical works~\cite{Hertzberg05,
  Rodrigues06:torque, Rodriguez07:PRL, Emig07:ratchet,
  RodriguezJo08:PRA, Rosa08:PRL, RahiZa10, LevinMc10} have predicted a
variety of situations in which these forces can be modified by using
complex microstructures.  In addition, by utilizing different choices
of materials the Casimir force can be changed in both
magnitude~\cite{Man09} and sign~\cite{Feiler08,
  Munday09,RodriguezMc10:PRL}.  However, with some
exceptions~\cite{RahiZa10, LevinMc10}, the qualitative aspects of
these effects can be explained through a combination or competition of
forces calculated using some form of the proximity force approximation
(PFA, a common heuristic description of the Casimir force as pairwise
interactions between parallel surface patches)~\cite{Derjaguin56}.  It
is therefore of interest to consider situations in which geometry not
only allows the Casimir force to be modulated (e.g., by reducing or
changing the sign of the force), but also creates effects that cannot
be accounted for by PFA.

\begin{figure}[tb]
\includegraphics[width=1.0\columnwidth]{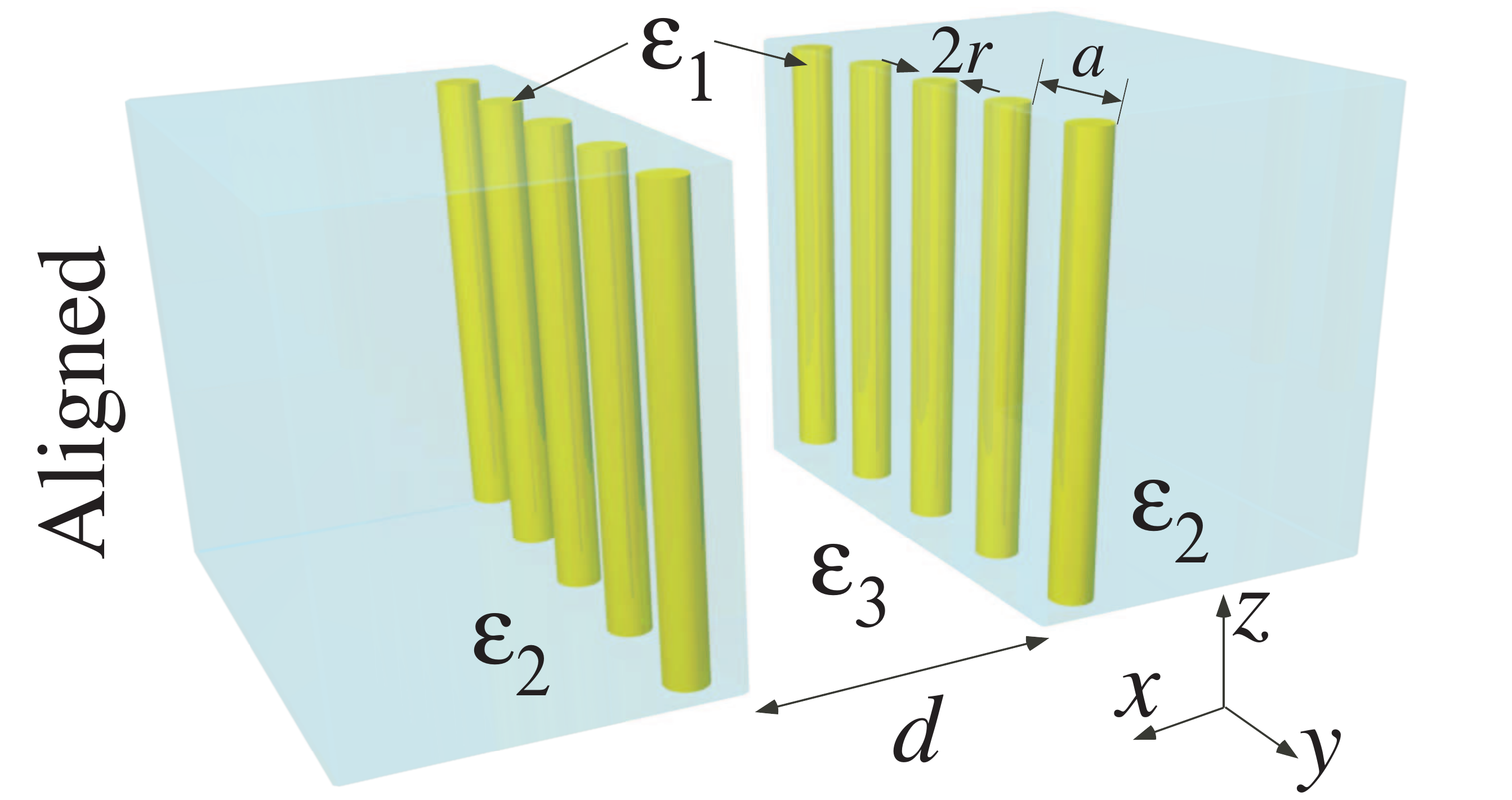}
\includegraphics[width=1.0\columnwidth]{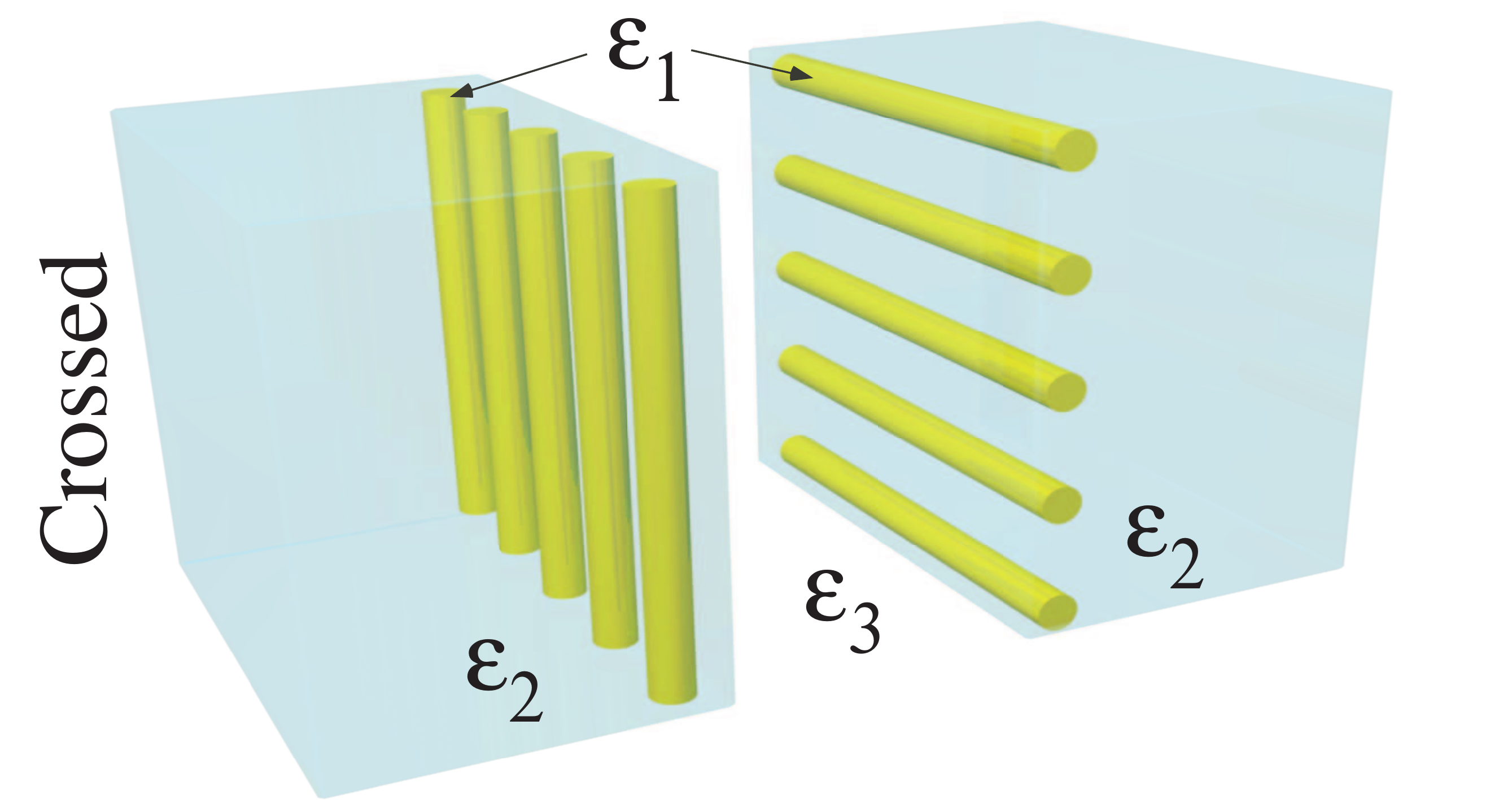}
\centering
\caption{Configuration examined in the text.  Each slab consists of a
  single periodic array of cylindrical wires embedded in a
  semi-infinite substrate; the goal is to create for each slab an
  effective anisotropic medium.  Of interest here is the change in the
  force from when the wires are aligned (Top) and when they are
  crossed (Bottom).  When the medium between the plates is a properly
  chosen fluid, we show that the force switches sign as a function of
  orientation for certain separations.}
\label{fig:config}
\end{figure}

In this article, we introduce and examine a geometry that exhibits
both of these qualities.  We examine the configuration shown
in~\figref{config}, consisting of two identical microstructured slabs
consisting of periodic arrays of dielectric nanowires.  We compute the
exact Casimir force for this geometry using a combination of existing
scattering theory techniques~\cite{Kushta00,Yasumoto04,Rahi09:PRD}, to
be described below.  We find that the microstructure of the slabs
leads to a number of interesting qualitative effects: the force
between the slabs can be dramatically modulated by rotating the two
slabs at fixed surface-surface separation.  In vacuum, the force
between the slabs can be significantly reduced (more than halved as
the rotation angle changes), and if the slabs are immersed in a fluid,
the sign of the force can flip.  It turns out that, even at moderate
separations $d/a \sim 1$, PFA (an uncontrolled approximation) cannot
capture either of these effects.  Specifically, consider the behavior
of the Casimir force between the slabs as they undergo lateral
translations $y$ (at $y=0$ the slabs are mirror-symmetric) and
rotations $\theta$.  At $\theta = 0$, the slab microstructures are
aligned as in~\figref{config} (upper), while for $\theta=\pi/2$ the
slabs are crossed, as in~\figref{config} (lower).  As $y$ and $\theta$
are varied, the slab surface-surface separation $d$ is kept fixed.
From a simple geometric argument, it is clear that for
vacuum-separated wires PFA predicts the following bound on the Casimir
force between two identical slabs:
\begin{equation}
F_{\mathrm{aligned}, y = a/2} \leq F_\mathrm{crossed} \leq
F_{\mathrm{aligned}, y = 0}
\label{eq:Intro_bound}
\end{equation}
(a positive force is attractive).  Here $F_\mathrm{crossed}$ denotes
the force when $\theta = \pi /2$.  This bound is insensitive to the
details of the exact PFA used, and simply relies on the fact that the
wire surface-surface separations are minimized at $\theta = 0, y = 0$
and maximized when $\theta = 0, y = a/2$.  Although a PFA prediction
must be valid as $d/a\rightarrow 0$, for the systems examined in this
work we find that the bound~\eqref{Intro_bound} is violated even at
moderate distances $d/a \sim 1$.  For vacuum-separated
metallic/dielectric (e.g., gold or silica) nanowires and no substrate,
this implies that changing the orientation of the wire arrays (i.e.,
their geometry) plays a stronger role in reducing the force than
simply reducing the pairwise surface-surface separation between the
slabs.  In another case to be discussed, in which the wires are gold,
the substrate silica, and the fluid ethanol, the
bound~\eqref{Intro_bound} is also valid.  However, we find from exact
calculation that while PFA predicts $F_{y = a/2} < 0$ and
$F_\mathrm{crossed} > 0$, the opposite is in fact true, i.e., aligned
slabs are always attracted to each other, while crossed slabs are
\emph{repelled}.  Therefore, in our case at $d/a\sim 1$ the sign of
the force can therefore be modulated by rotating the slabs (but not
translating them), an interesting possibility for experiments that we
discuss later.

To understand these effects qualitatively, we find that a much more
suitable framework is the effective medium approximation (EMA), in
which the microstructured slabs of~\figref{config} are describable as
an homogeneous medium with a given anisotropic permittivity tensor
$\stackrel{\leftrightarrow}{\varepsilon} = \mathrm{diag}\left(
\varepsilon_{||}, \varepsilon_\perp, \varepsilon_\perp\right)$ (here
and below, $\stackrel{\leftrightarrow}{\varepsilon}$,
$\varepsilon_{||}$ and $\varepsilon_\perp$ will be used to denote
effective, homogenized permittivities).  We are able to compute the
effective $\varepsilon$ tensor for this configuration from our
scattering method~\cite{Rosa08}, and we find that for the
gold-silica-ethanol configuration, the deduced effective-medium
parameters follow an ascending sequence $\varepsilon_\perp <
\varepsilon_\mathrm{fluid} < \varepsilon_{||}$.  This ascending
sequence is known to lead to repulsion~\cite{Dzyaloshinskii61} for
uniform, isotropic materials, and we find below that the effect is
also present for anisotropic materials and individual polarizations
(although in this case, as discussed below, an ascending sequence is
not sufficient for repulsion).  One expects the EMA to be strictly
valid as $d\rightarrow\infty$, however in our case it turns out,
somewhat surprisingly, to be qualitatively accurate down to $d/a\lesssim
1$ (a mathematical proof of why the EMA holds at such short
separations, at least for crossed slabs, is given in~\Appref{appB}).
Further, we find that for realistic materials (gold wires, silica
substrate, immersed in an ethanol fluid) and geometry parameters, this
orientation induced repulsion at fixed surface-surface separation
between the two slabs holds down to separations $d/a \sim 0.5$.

Previous works~\cite{Rosa08, Zhao09} have discussed the possibility of
using geometry to create artificial constitutive relations (e.g.,
effective permeability $|\stackrel{\leftrightarrow}{\mu}| >
|\stackrel{\leftrightarrow}{\varepsilon}|$ [under a suitable matrix
norm] to obtain Casimir repulsion or chirality $\kappa \neq 0$ to
achieve force reduction).  However, it has been rigorously
proven~\cite{Rahi10:PRL} that no such media with
$\stackrel{\leftrightarrow}{\mu} >
\stackrel{\leftrightarrow}{\varepsilon}$ can be constructed from
metallic/dielectric constituents so as to exhibit repulsion in vacuum,
and computations involving the exact
microstructures~\cite{McCauleyZh10} have shown that chirality effects
are only present at separations so large that they cannot conceivably
be detected.  Therefore, while in some circumstances EMA can be a
useful qualitative guide (and rigorously accurate in certain limits),
it must be used with caution---ideally, as a supplement to exact
calculations.  Orientation dependence (and the resulting Casimir
torque) between slabs has previously been considered for two
birefringent plates with weak anisotropy~\cite{Enk95:torque, Shao05,
  Munday05}, and for corrugated metallic
plates~\cite{Rodrigues06:torque}.  Although the torques in these
systems are in principle measurable, the change in the force with
orientation is small, and the forces are always attractive.
\citeasnoun{Kenneth00:torque} showed a large orientation dependence
for theoretical uniaxial conductors, and suggested a possible
realization via nanowire arrays, but without calculation in the latter
case and without changing the sign of the force.

In the present work, we analyze this effect for
periodically-patterned, vacuum-separated suspended membranes (which
form a potentially promising medium for Casimir force
measurements~\cite{Rodriguez11:OL}) and for gold wires on silica
substrate, immersed in ethanol.  In both cases, we
find~\eqref{Intro_bound} is violated down to $d/a \sim 1$.  In the
former case, we find that the force is $70\%$ lower for crossed slabs
compared to aligned slabs (for any $y$) at large separations, while at
separations comparable to the unit cell (e.g., $100\,\mathrm{nm}$)
there is a more modest, but still significant, $30\%$ reduction.  For
the latter system, we find that aligned slabs are attractive for all
$y$ down to $d/a \sim 0.5$ while crossed slabs are repulsive in this
range.  These examples demonstrate a system in which effective-medium
theory is correct and gives predictions that differ from PFA in a
highly nontrivial way.  Given that nanowire arrays below
$15\,\mathrm{nm}$ can be fabricated with current
technology~\cite{Morecroft2009} and that for these dimensions the
predicted effects occur down to sub-$100\,\mathrm{nm}$ length-scales,
these effects should be experimentally accessible.  After presenting
these results, we argue how they may be experimentally detectable
(assuming suitable fabrication techniques) and estimate force
magnitudes in the hypothetical case where one of the slabs is replaced
by either a sphere or cylinder with a wire pattern stamped on its
surface.

\section{Method}
\label{sec:Method}

In this work, we perform Casimir force calculations with a
semi-analytic scattering method using a combination of results from
\citeasnoun{Rahi09:PRD,Kushta00,Yasumoto04}, which efficiently
computes the exact Casimir force between periodic arrays with one
dimensional of translation-invariance.  Our implementation differs
somewhat from previous scattering methods
(e.g.,~\citeasnouns{Lambrecht09, Davids10}) in that it is particularly
well-suited to unit cells with objects of compact cross-section, such
as the wires in~\figref{config} (rather than, e.g., extended
rectangular gratings). While we are primarily interested in objects of
circular cross-section in the current work, the present method can be
extended in a straightforward manner to treat unit cells with
arbitrarily shaped compact objects using an existing boundary-element
method~\cite{ReidRo09, Reid11:2D}.  In addition, we have checked the
results with a brute-force finite-difference time-domain
(FDTD)~\cite{RodriguezMc09:PRA, McCauleyRo10:PRA} method and found
good agreement, and additionally these computations show similar
results for square wires.  Also, the effects are not significantly
different at zero temperature and 300 K, so we work exclusively in the
former limit.

The two plates are separated from each other by a distance $d$ in the
$x$ direction, and the slab is termed $y$-directed if the wires are
along the $y$ axis (similarly for $z$).  The slabs are aligned if both
slabs are $y$- or $z$-directed, and crossed if they are not.  The
zero-temperature Casimir interaction energy for a $q$-directed slab
displaced by $\vec{x}$ from a $q^\prime$ directed slab ($q,q^\prime =
y,z$) is:
\begin{equation}
E = \frac{\hbar c}{2\pi} \int_0^\infty d\xi \int
\frac{d^2\vec{k}_{yz}}{4\pi^2}\log \det \left(\mathbb{I} -
\mathbb{R}_{1}^{(q^\prime)} \mathbb{U}^\dagger
\mathbb{R}_{2}^{(q)} \mathbb{U} \right)
\label{eq:energy}
\end{equation}
where $\mathbb{R}_{1,2}$ are the scattering matrices of planewaves for
the two slabs, $\mathbb{U}(\vec{x})$ is the planewave translation
matrix for relative displacement $\vec{x}$ between the slabs, and the
integral of transverse wavevector components $k_y,k_z$ is over the
first Brillouin zone.  See~\citeasnoun{Rahi09:PRD} for a detailed
derivation and a partial review of precursors~\cite{Emig07, Kenneth08,
  Neto08}.  For the present work, we require an efficient method of
computing the scattering matrices from periodic arrays, which we
describe in more detail in~\Appref{Method-scattering}.

For sufficiently large distances, the relevant frequencies and
wave-vectors are unable to probe the structure of the arrays and
consequently an effective medium approximation should produce good
results. In our case, such an effective medium should have a much
larger conductivity in the direction of the wires as compared to the
other orthogonal directions even in the static limit (this being one
of the motivations for the current configuration), so it is clear that
an anisotropic EMA is called for.  Fortunately, Casimir interactions
between anisotropic homogeneous media have been studied by several
authors~\cite{Parsegian72, Barash75, Barash78, Philbin08, Rosa08},
allowing us to build upon their results.  We use the method
of~\citeasnoun{Rosa08} (outlined in~\Appref{Method-EMA}) to obtain the
scattering matrices assuming a known permittivity
$\stackrel{\leftrightarrow}{\varepsilon}$.  In the subsequent
analysis, we also require the inverse procedure: given the scattering
matrices (computed using the method of~\Appref{Method-scattering}) of
the exact structure, retrieve the best-fit EMA
$\stackrel{\leftrightarrow}{\varepsilon}$.  This procedure quickly
becomes intractable if arbitrary
$\stackrel{\leftrightarrow}{\varepsilon}$ and
$\stackrel{\leftrightarrow}{\mu}$ are allowed.  Instead, we assume
$\mu_{ij} = \delta_{ij}$ and
$\stackrel{\leftrightarrow}{\varepsilon}_{xx} =
\stackrel{\leftrightarrow}{\varepsilon}_{yy}$ exhibit no dependence on
$\vec{k}_\perp$ aside from its polarization.  For
$\left|\vec{k}_\perp\right| \rightarrow 0$ the scattering matrices for
each polarization reduce to the standard Fresnel formula for
reflection off isotropic interfaces.  Their inversion then yields the
effective dielectric tensor:

\begin{equation}
\varepsilon_{||}(i\xi) = \varepsilon_\mathrm{fluid} \left(\frac{ 1 - R_{||}}{1 + R_{||}}\right)^2,
~~\varepsilon_\perp(i\xi) = \varepsilon_\mathrm{fluid} \left(\frac{1 + R_\perp}{1 - R_\perp}\right)^2
\label{eq:retrieval}
\end{equation}

\section{Results}
\label{sec:Results}

In the configuration of~\figref{config}, the wires have radius
$r=0.3\,a$ and period $a$; we take the wire centers to be in the
substrate and the wire surface to be tangent to the substrate surface
to maximize the wire-wire interactions.  The permittivities of wires,
substrate, and fluid are respectively $\varepsilon_1(i\xi)$,
$\varepsilon_2(i\xi)$ and $\varepsilon_3(i\xi)$ for imaginary
frequency $\xi$.  The materials used for the wires, substrate, and
fluid are gold, ethanol, and silica, respectively.  For gold, we use a
plasma model with $\xi_p = 1.36\times 10^{16}\,\mathrm{rad/sec}$ (the
addition of a small loss term does not change the results
significantly).  For silica and ethanol we use standard oscillator
models~\cite{Milling96, Bergstrom97}.

\subsection{Vacuum-separated slabs}
\label{sec:Results-vac}

\begin{figure}[tb]
\includegraphics[width=1.0\columnwidth]{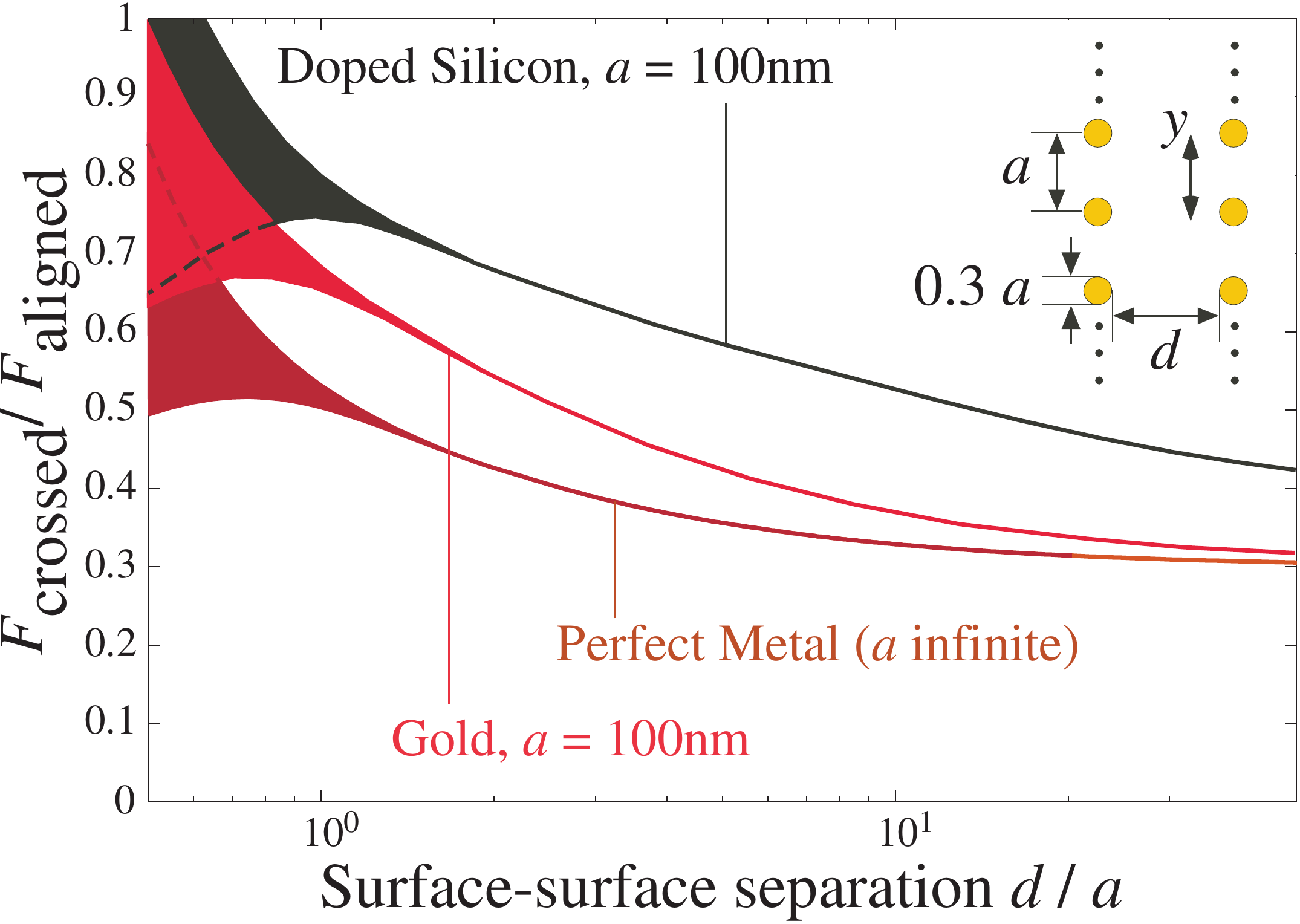} \centering
\caption{Ratio of the Casimir force between crossed wire arrays
  $(F_\mathrm{crossed})$ over aligned wires $(F_\mathrm{aligned}(y))$
  between vacuum-separated nanowires of perfect metal, gold, and doped
  silicon (dopant density $10^{20}/\mathrm{cm}^3$).  $a \rightarrow
  \infty$ corresponds to using non-dispersive
  $\varepsilon(i\xi\rightarrow 0)$ for all materials (because $\xi$ is
  in units of $c/a$).  Shaded regions indicate the range of values
  that $F_\mathrm{aligned}$ takes over all relative lateral
  displacements $y$ of the wire centers, with their higher (lower)
  boundary indicating the force for $y = 0$ ($y = a/2)$.  PFA
  predicts, via~\eqref{Intro_bound}, that the maximum of this ratio
  should exceed 1, which is only true for Si at $d/a \sim 0.6$, and
  for gold/perfect metal at lower separations.  Inset: details of the
  configuration examined.}
\label{fig:results-vac}
\end{figure}

We first compute the Casimir forces when the intervening medium is
vacuum ($\varepsilon_3=1$), comparing the force
$F_\mathrm{aligned}(y)$ for aligned slabs with the force
$F_\mathrm{crossed}$ for crossed slabs.  In the aligned case, the
force can depend on $y$ (leading to a lateral component of the force),
and in this case $F_\mathrm{aligned}$ refers only to the \emph{normal}
component of this force.  In the crossed case $\theta = \pi/2$ there
is no $y$-dependence.  Although there is no sign change for
vacuum-separated slabs, this configuration is of interest because it
can be fabricated as a single suspended-membrane structure for each
orientation~\cite{Rodriguez11:OL}, and may be easier to work with than
a fluid system.  \Figref{results-vac} shows results for wires composed
of perfect metal, gold, and heavily doped silicon, the latter being
more conventional for fabrication.  We plot the ratio
$F_\mathrm{crossed}/F_\mathrm{aligned}(y)$ (the shaded regions
indicate the full range of this ratio as $y$ is varied), which serves
two purposes.  First, it indicates the required relative accuracy in a
force measurement needed to discern the orientation-dependence of the
force in an experiment.  Second, it indicates the transition from the
PFA to the EMA regimes via~\eqref{Intro_bound}: if the force is
determined as a sum of pairwise interactions, then the total force is
maximized when the pairwise surface-surface distances are minimized.
A simple geometric argument shows that the net distance is maximized
for $\theta = 0, y = a/2$.  On the other hand, this distance is
minimized for $\theta = 0, y = 0$.  The bound~\eqref{Intro_bound} then
follows.  It turns out that the force is always attractive in this
situation, so we have the further bound $F_\mathrm{aligned}(y = a/2) >
0$.  Therefore, PFA predicts that there is a range of $y$ such that
$F_\mathrm{crossed}(y) / F_\mathrm{aligned} > 1$.  By contrast, EMA
predicts the inequality $F_\mathrm{crossed} < F_\mathrm{aligned}$
(there is no $y$-dependence in this approximation).  This inequality
stems from the following scattering-theory argument: for $d/a \gg 1$,
the exponential suppression of $k_y, k_z \neq 0$ in
$\mathbb{U}$~\cite{Rahi09:PRD} implies that the force is dominated by
the scattering of planewaves at normal incidence ($k_y=k_z=0$).  For
normal incidence, the reflection matrix $\mathbb{R}^{(z)}$ is
anisotropic but diagonal in polarization, and can be computed from an
effective anisotropic dielectric tensor:
$\stackrel{\leftrightarrow}{\varepsilon}(i\xi)$:
\begin{equation}
\stackrel{\leftrightarrow}{\varepsilon}(i\xi) = \mathrm{diag}\left( \varepsilon_{xx},\varepsilon_{yy},\varepsilon_{zz}\right) \Rightarrow \mathbb{R} \approx \mathrm{diag}\left(R_{||}, R_\perp\right)
\end{equation}
where $R_{||}$ and $R_\perp$ are the matrix elements for incident
light polarized parallel and perpendicular, respectively, to the
wires.  For normal incidence, $\varepsilon_{xx}$ does not contribute,
and we define $\varepsilon_{||}=\varepsilon_{zz}$ and
$\varepsilon_\perp=\varepsilon_{yy}$.  Due to the high permittivity of
the wires and the low permittivity of the substrate, we expect
$|R_\perp| \ll |R_{||}|$ and $R_\perp R_{||} \geq 0$ [note that
  $R(i\xi)$ is real].  When the wires are aligned, the relevant
product of scattering matrices in the energy integrand is $\left(
R_{||}^2 + R_\perp^2 \right) \approx R_{||}^2$, and when they are
crossed the term is $2 R_{||} R_\perp$, implying $F_\mathrm{crossed} <
F_\mathrm{aligned}$.  From~\figref{results-vac}, this bound is clearly
violated in doped Si for $d / a < 0.6$, and for other materials at
smaller $d$.  Note that this includes a region $d/a < 2$ for which
there is still a strong $y$-dependence of this ratio, indicating that
while EMA is not strictly valid, it still has more predictive power
than PFA.

As $d/a \rightarrow \infty$, the force ratios should approach a
constant determined by
$\stackrel{\leftrightarrow}{\varepsilon}(\xi\rightarrow 0)$: this
ratio is $\approx 0.29$ for all three materials.  However, this limit
is approached very slowly ($O(d^{-1/2})$ for perfect metal/gold
wires).  The EMA $d$-dependence is due both to constituent material
dispersion and an effective geometric dispersion in
$\stackrel{\leftrightarrow}{\varepsilon}(i\xi)$ induced by the
geometry of the wires.  As $\xi\rightarrow 0$, $\varepsilon_{||}
\rightarrow \varepsilon_1(i\xi)$ and $\varepsilon_\perp \rightarrow
\varepsilon_2(i\xi)$.  The former limit follows from the $1/\xi^2$
divergence of $\varepsilon_1$ as $\xi\rightarrow 0$ (if we add a
dissipative term to $\varepsilon_1$, the divergence is only $1/\xi$
but this crossover occurs at frequencies too low to affect the present
results).  The latter limit follows from the fact that the static
polarizability of the wires in the transverse direction is finite,
implying that their contribution to the scattering amplitude vanishes
as $\xi \rightarrow 0$.  Therefore, as $\xi\rightarrow 0$,
$\varepsilon_\perp$ depends only on the substrate, which has
semi-infinite extent; we have checked this relation numerically and
found good agreement.  $\lim_{d\rightarrow
  \infty}F_\mathrm{crossed}/F_\mathrm{aligned}$ can then be obtained
by computing the force between two anisotropic plates, with
$\varepsilon_{||} = \infty$ and $\varepsilon_\perp =
\varepsilon_2(0)$.

For all $d/a \gg 1.5$, the orientation-dependence of the force is
quite strong ($\gg 30\%$).  In this range, the absolute pressure is
approximately $2\%$ of the corresponding pressure between two
homogeneous perfect metal plates.  As differences of this magnitude
between vacuum-separated plates have been measured on the
$100-\,\mathrm{nm}$ length scale, we are hopeful that this departure
from PFA can be detected experimentally.

\subsection{Fluid-separated slabs and tunable repulsion}
\label{sec:Results-fluid}

In the previous example the force is always attractive, because
$\varepsilon_1(i\xi), \varepsilon_2(i\xi) \geq \varepsilon_3(i\xi)$
for all $\xi$~\cite{Rahi10:PRL}.  However, if the medium between the
slabs is a fluid such that $\varepsilon_1(i\xi) > \varepsilon_3(i\xi)
> \varepsilon_2(i\xi)$, the above EMA analysis predicts that the force
in the crossed configuration should be repulsive as $d/a\rightarrow
\infty$ since by the Fresnel formula at $\xi = 0$, $R_\perp R_{||} <
0$, leading to the well-known Casimir repulsion for an ascending
sequence of $\varepsilon$ for each polarization~\cite{Munday09}. By
contrast, in the aligned configuration the force is attractive when
$y=0$ due to mirror symmetry~\cite{KennethKl06}, and the intuition
that this attraction holds for non-zero $y$ is confirmed by
calculation.  However, strong corrections to this argument occur at
finite separations: waves with nonzero $k_y$ and $k_z$ contribute, and
are not in general polarized along the $y$ or $z$ axes.  These waves
will couple to both $\varepsilon_{||}$ and $\varepsilon_\perp$ on each
reflection.  The sign of the resultant reflection coefficient will
usually be the same for both $z$- and $y$-directed slabs, leading to
attractive contributions.  Further, the microstructure can also lead
to significant corrections, possibly eliminating the effect for
separations comparable to the unit cell $a$~\cite{McCauleyZh10}.  As
such separations are necessarily required for experiments, we require
exact results to verify that this effect persists in experimentally
accessible regions.

\begin{figure}[tb]
\includegraphics[width=1.0\columnwidth]{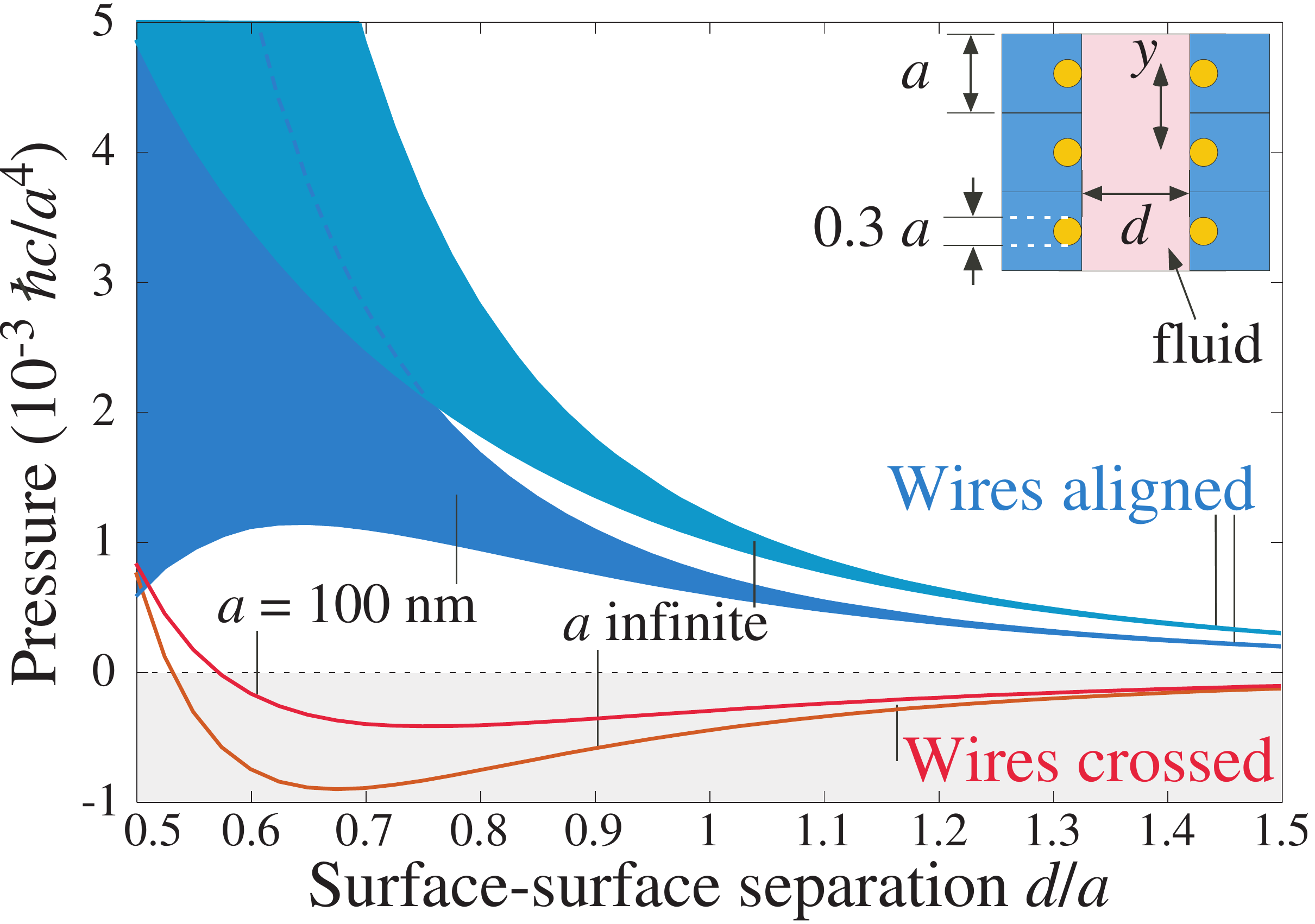} 
\centering
\caption{Casimir pressure for the same parameters as the inset
  to~\figref{results-vac}, with a silica substrate and ethanol between
  the two slabs.  Positive values indicate an attractive force,
  negative values (shaded) repulsive.  Both $a\rightarrow\infty$ and
  $a=100\,\mathrm{nm}$ are shown.  As in~\figref{results-vac}, shaded
  regions denote the range of $y$-displacements.  When the wires are
  aligned (blue), the force is always attractive, but when the wires
  are crossed (red), the force turns repulsive at a critical
  separation $d > d_c$, which depends on $a$.}
\label{fig:results}
\end{figure}

The exact results are shown in~\figref{results}; as
in~\figref{results-vac}, shaded regions show the $y$-dependence. The
results show a clear attractive-repulsive transition as the slab
orientation is varied between the aligned and crossed configurations.
This effect persists for both the ideal case of perfect metals
$(a\rightarrow \infty)$ and dispersive materials at
$a=100\,\mathrm{nm}$; both show qualitatively similar behavior.  For
crossed slabs, the repulsion is fairly flat over a $\sim
40\,\mathrm{nm}$ range.  At first sight, it is temping to ascribe the
repulsive force observed for crossed wires to PFA-line interactions
between opposing areas where metallic wires face dielectric substrate
and thus feel a repulsive force.  However, if this were the case, then
by the same argument (and~\eqref{Intro_bound}), $F_\mathrm{aligned}(y
= a/2)$ should exhibit an even stronger repulsion.  This is clearly
not the case in~\figref{results} for $d/a > 0.5$; rather, the results
for repulsion are consistent with the EMA argument given above.  To
estimate the magnitude of this repulsion, for $a=100\,\mathrm{nm}$ the
repulsive pressure is approximately $1\%$ of the pressure between
parallel perfectly-conducting plates separated by a comparable
distance $d$ in vacuum, e.g, $\sim
0.13\,\mathrm{pN}/\mu\mathrm{m}^2$.  Repulsive forces in fluids on
this order of magnitude have been measured~\cite{Munday09}, therefore
these forces are potentially within reach of current or near-future
measurement techniques (the force in more realistic sphere-plate and
cylinder-plate geometries will be considered in~\secref{geometries}
below).  Interestingly, when the wires are crossed and $d$ varies
there is an attractive-repulsive transition at a critical separation
$d = d_c$.  This leads to an unstable equilibrium with respect to
$d$. (The crossed configuration is always unstable with respect to
orientation.)  This transition, as mentioned above, is attributed to
the eventual dominance of attractive forces as separation goes to
zero.  However, we will see below that such a transition is predicted
by EMA as well, and is therefore not due entirely to proximity
effects.

\begin{figure}[tb]
\includegraphics[width=1.0\columnwidth]{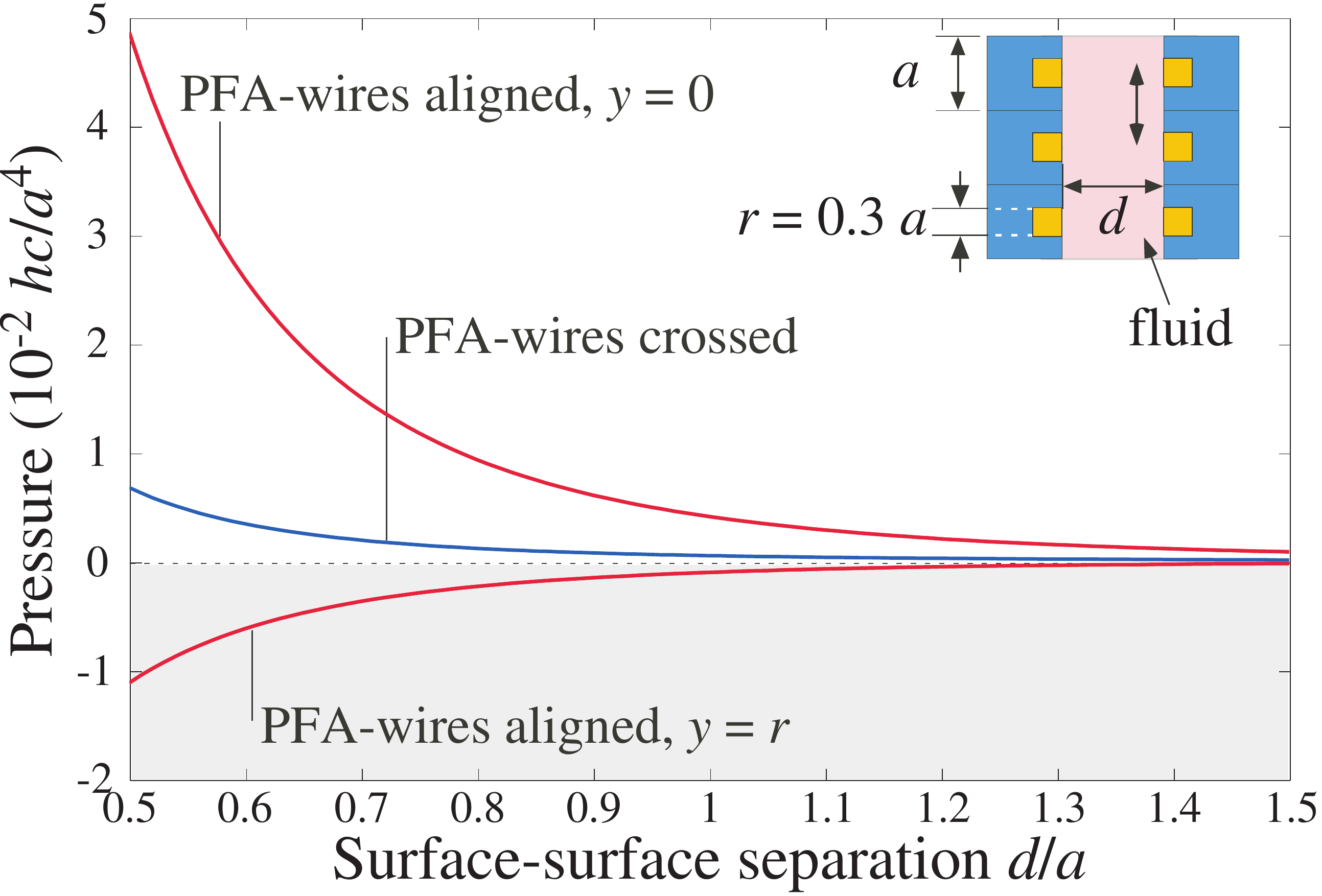}
\centering
\caption{Forces computed for the configuration of~\figref{results}
  using the Proximity Force Approximation (PFA), for $a =
  100\,\mathrm{nm}$.  For convenience in applying the PFA, square
  wires of width $r$ are used instead of circular wires (the exact
  results are not qualitatively changed).  We see that the behavior
  of~\figref{results} is not qualitatively captured: the force is
  always attractive for crossed wires, and the
  bound~\eqref{Intro_bound} is strictly satisfied.  Similar results
  are obtained for $a\rightarrow\infty$.}
\label{fig:PFA-fig}
\end{figure}

Before continuing, it is interesting to see what the PFA prediction
for the force in this system is.  We simplify matters by computing the
PFA assuming instead wires of square $r \times r$ cross-section, with
centers a distance $r/2$ beneath the substrate surface (by minimizing
the amount of curved surface, we expect this to maximize the range of
separations for which PFA is accurate).  Using FDTD
computations~\cite{RodriguezMc09:PRA, McCauleyRo10:PRA}, we have found
that this system exhibits behavior qualitatively similar to that
of~\figref{results}.  The results for $a = 100\,\mathrm{nm}$ are shown
in~\figref{PFA-fig} (taking $a \rightarrow \infty$ yields a
qualitatively similar curve).  Inspection reveals that the PFA
prediction exhibits the \emph{opposite} behavior for the sign of the
force as the full numerical calculations: the PFA force is exclusively
\emph{attractive} for crossed wires, but for aligned wires shifted by
$y = a/2$ the force becomes \emph{repulsive}.

Although PFA fails qualitatively and quantitatively to predict the
results of~\figref{results-vac} and~\figref{results}, we have not yet
examined the quantitative accuracy of EMA.  In the next section, we
will examine a description of the force in terms of the EMA,
rigorously valid in the regime ($d/a \rightarrow \infty$).  We will
see that EMA gives qualitatively correct predictions for the magnitude
and sign of the forces down to separations comparable to the unit cell
size.

\subsection{Comparison with EMA}
\label{sec:Results-EMA}

\begin{figure}[tb]
\includegraphics[width=1.0\columnwidth]{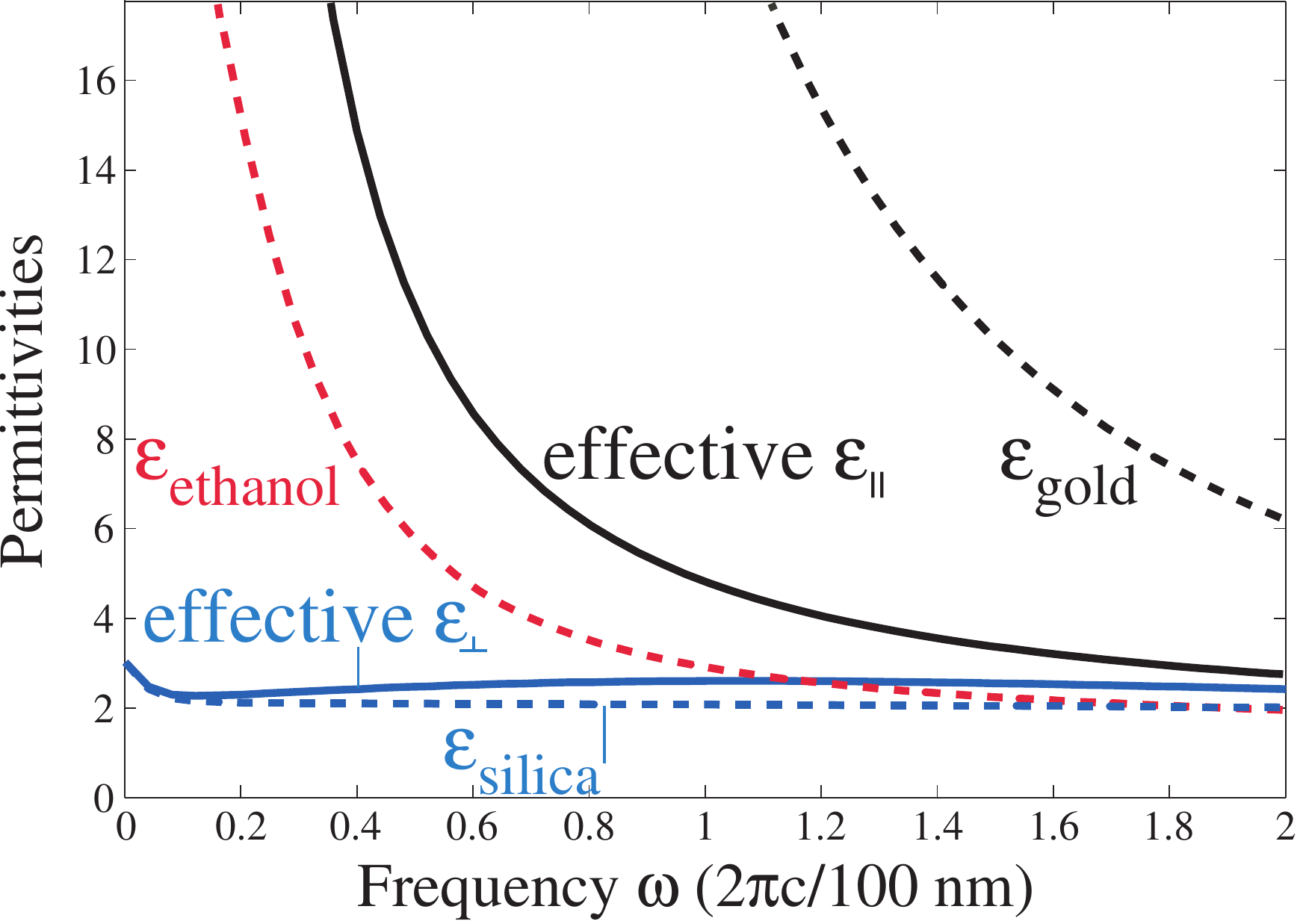} 
\centering
\caption{Effective-medium parameters for the system
  of~\figref{results} derived from scattering data for a length scale
  $a = 100\,\mathrm{nm}$.  $\varepsilon_{||}$ is the effective
  dielectric parallel to the wires (black), and $\varepsilon_\perp$
  the effective dielectric perpendicular to the wires (blue).  The
  dielectrics of the constituent materials gold, silica, and ethanol
  are shown for reference.  These dispersions will be used to compute
  the Casimir force in the effective-medium approximation below.}
\label{fig:effective-params}
\end{figure}

\begin{figure}[tb]
\includegraphics[width=1.0\columnwidth]{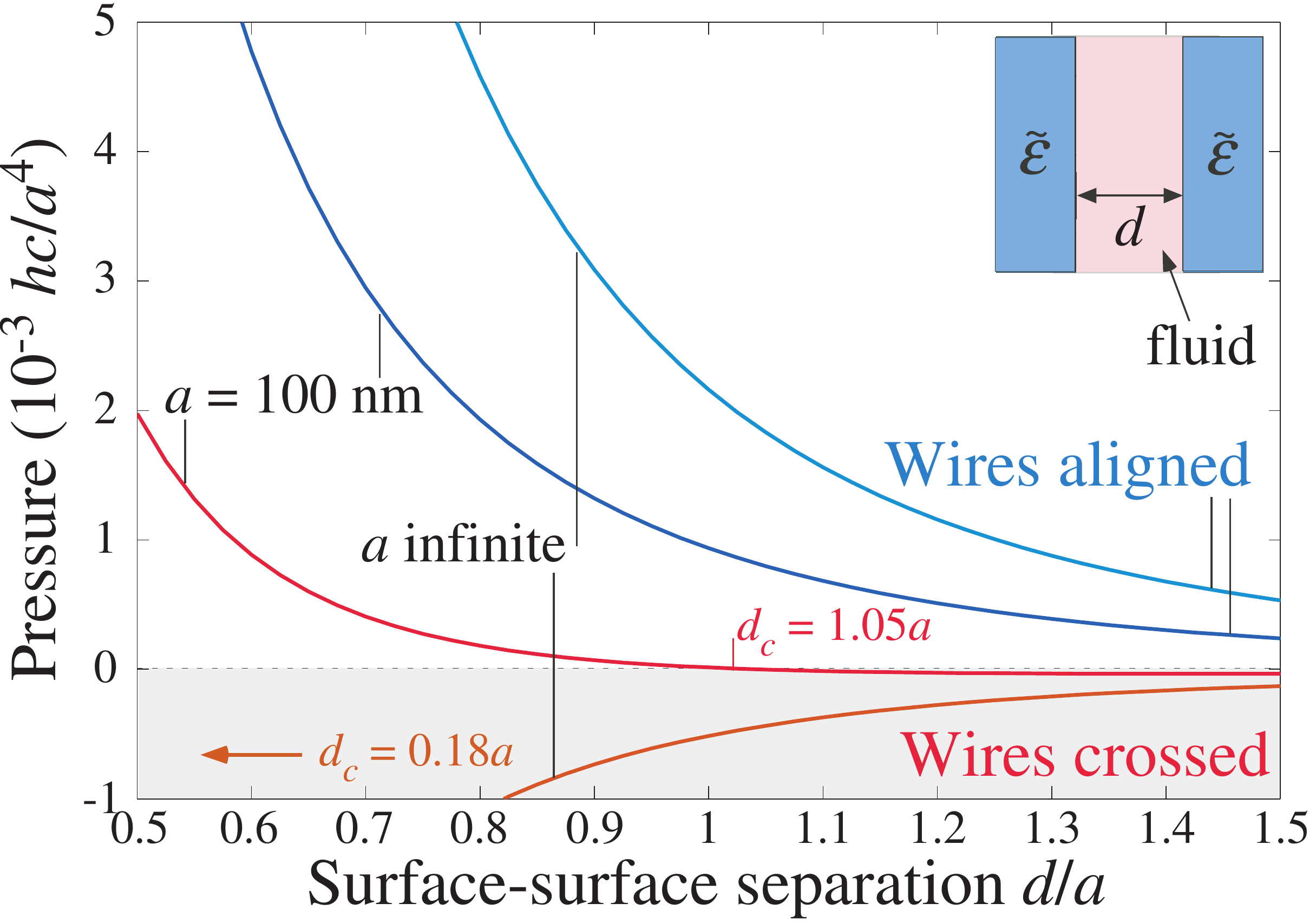}
\centering
\caption{Results for the force computed in the Effective-medium
  Approximation (EMA) using an effective homogeneous, anisotropic
  permittivity tensor $\stackrel{\leftrightarrow}{\varepsilon}$, via the method described
  in~\Appref{Method-EMA}.  As in~\figref{results}, results for both
  $a$ infinite (i.e., non-dispersive materials) and $a =
  100\,\mathrm{nm}$ are shown, for both crossed and aligned
  configurations.  Although quantitatively inaccurate over the range
  plotted, the EMA does qualitatively capture the behavior of the
  force a opposed to PFA.  Also shown are $d_c$, the location of the
  attractive-repulsive transitions for the crossed wires ($d_c =
  0.18a$ for $a \rightarrow \infty$ and is not shown).}
\label{fig:EMA-force}
\end{figure}

In this section, we examine the extent to which the EMA analysis given
above in~\secref{Results-vac} predicts the correct results, using the
method of~\Appref{Method-EMA}.  We use a simplified EMA, assuming that
  $\stackrel{\leftrightarrow}{\mu}(i\xi) = 1$ and
  $\stackrel{\leftrightarrow}{\varepsilon}(i\xi)$ is isotropic in the
  plane perpendicular to the wires.  In this case, $\varepsilon_{||}$
  and $\varepsilon_\perp$ are the only EMA parameters involved; these
  are retrieved by applying~\eqref{retrieval} to the scattering
  matrices computed from the exact structure.  The EMA-predicted force
  is then computed from the method of~\Appref{Method-EMA}.  A plot of
    the retrieved parameters for $a=100\mathrm{nm}$ is shown
    in~\figref{effective-params}.  Shown for reference
    in~\figref{effective-params} are $\varepsilon(i\xi)$ of the
    constituent materials ethanol, silica, and gold.  As expected,
    $\varepsilon_\perp$ approaches $\varepsilon_\mathrm{silica}$ for
    low $\xi$, while for higher $\xi$ the wires increase the
    effective permittivity.  In $\varepsilon_{||}$, we see a significant
    geometric dispersion: in fact, at low $\xi$,
    $\varepsilon_{||}(i\xi) \sim 1 + \left(\xi_p^\prime/\xi\right)^2$,
    where $\xi_p^\prime \sim 0.36 \,\xi_p$ is a new, effective plasma
    frequency for the gold wires.  This is similar to an argument
    presented in~\citeasnoun{Pendry96}, where the effective
    dielectrics of square arrays of wires along the wire axis can be
    described by a plasma model with a reduced plasma frequency (as we
    have only included a single row of wires,~\citeasnoun{Pendry96}'s
    result cannot be directly applied, but its basic idea remains).  A
    similar effect holds when the wires are perfect conductors, where
    $\xi_p^\prime = 7.9$.  This accounts for the $d$-dependence in the
    EMA regime of~\figref{results-vac}: the geometry introduces an
    effective length scale into the system, given by $\xi_p$.

We use the retrieved $\varepsilon_{||}$ and $\varepsilon_\perp$ to
compute $\mathbb{R}$ and hence the Casimir force for the
fluid-separated geometry.  Computation of the force with the EMA
parameters for the vacuum-separated slabs in the limit $d\rightarrow
\infty$ gives agreement with the results of~\figref{results-vac}.  The
results for the fluid-separated case are shown in~\figref{EMA-force}.
We find that, as opposed to PFA, EMA gives qualitatively accurate
(i.e., the same order of magnitude) predictions for both the magnitude
and sign of the force.  In particular, EMA predicts
attractive-repulsive transitions at some $d=d_c$, where for $d < d_c$
the force for crossed slabs is attractive.  The fact that
$\varepsilon_{||}(\xi)$ is a rapidly decreasing function of $\xi$,
while $\varepsilon_\perp$ increases, indicates that within the EMA the
force should receive attractive contributions from higher $\xi$ (which
dominate at small separations).  Further, the decrease of both
$\varepsilon_{||}$ and $\varepsilon_2$ with decreasing $a$ also
contributes to a reduction in repulsion.  This explains why the
repulsion for $a=100\,\mathrm{nm}$ is lower than for $a\rightarrow
\infty$.  With our simplified EMA the predicted values of $d_c$ are
very inaccurate: $d_c \approx \,0.18\,a$ for $a=\infty$ and $d_c
\approx 1.05\,a$ when $a=100\,\mathrm{nm}$.  However, the
$y$-independent regimes of Figs.~2--3 suggests that some EMA must be
valid in those regimes, and in this case a more accurate EMA would
involve both $\varepsilon_{xx} \neq \varepsilon_\perp$ as well as an
anisotropic effective $\stackrel{\leftrightarrow}{\mu}$.  A more
general model of dispersion would allow for a translation-invariant
but $\vec{k}$-dependent permittivity.  Fitting an effective
$\stackrel{\leftrightarrow}{\varepsilon}(\vec{k})$ is somewhat
complicated in this framework; rather, to explore the validity of this
``specular'' approximation we compute the force with all non-specular
(i.e., terms not conserving $\vec{k}_\perp + \vec{G}_\perp$, where
$\vec{G_\perp}$ is a reciprocal lattice vector) terms in the
scattering matrix $\mathbb{R}$ removed.  We find the surprising result
that, while this specular approximation does not give significantly
different predictions for the force for aligned wires, it gives much
more accurate predictions for $d_c$ and the magnitudes of the force
for crossed wires.  This indicating that non-specular scattering
events are suppressed when the wires are crossed, even at relatively
small separations $d/a \sim 0.5$.  We give a rigorous proof of this
result in~\Appref{appB}, using a recently-developed diagrammatic
expansion for the Casimir energy~\cite{Maghrebi10:arxiv}.

\subsection{Forces for other geometries}
\label{sec:geometries}

Parallel-plate configurations involving suspended membrane structures
show potential for new sets of experimental Casimir force measurements
for vacuum-separated geometries~\cite{Rodriguez11:OL}; such
measurements detect the Casimir force (or force gradient) through a
shift in either the optical spectrum or the resonance frequency of the
upper membrane.  However, to measure the sign change in the force
predicted here a measurement between fluid-separated objects must be
performed, where fluid damping is prohibitive.  Instead, a force
measurement involving an object (e.g., a sphere) mounted on an atomic
force microscope (AFM) tip~\cite{Munday09} is more realistic.  In this
case, alignment issues favor the use of one spherical and one planar
object~\cite{Lamoreaux97,moh1}, rather than the two planes considered
here.  For our purpose, a pattern of wires similar to~\figref{config}
would be need to be stamped on the surface of the object.  In this
section, we give force predictions both sphere-plate and
cylinder-plate geometries.  The latter case has more difficult
alignment issues, but still simpler than plate-plate alignment, and in
this case the Casimir force is much larger than for sphere-plate
setups~\cite{Brown-Hayes05}.  For both cases, the radius of curvature
$R$ is many orders of magnitude larger than the surface-surface
separation $d$ (e.g., $R = 200\,\mu\mathrm{m}$ and $d \sim
100\,\mathrm{nm}$).  Because $d, a \ll R$, it is appropriate to use a
hybrid PFA/exact method in which each unit of surface on the
sphere/cylinder feels the \emph{exact} pressure (as computed
in~\secref{Results-fluid}) between a plate-plate configuration of the
same surface-surface separation.  The result is asymptotically exact
in the limit $R \rightarrow \infty$ for fixed $d$ and $a$.

\begin{figure}[tb]
\includegraphics[width=1.0\columnwidth]{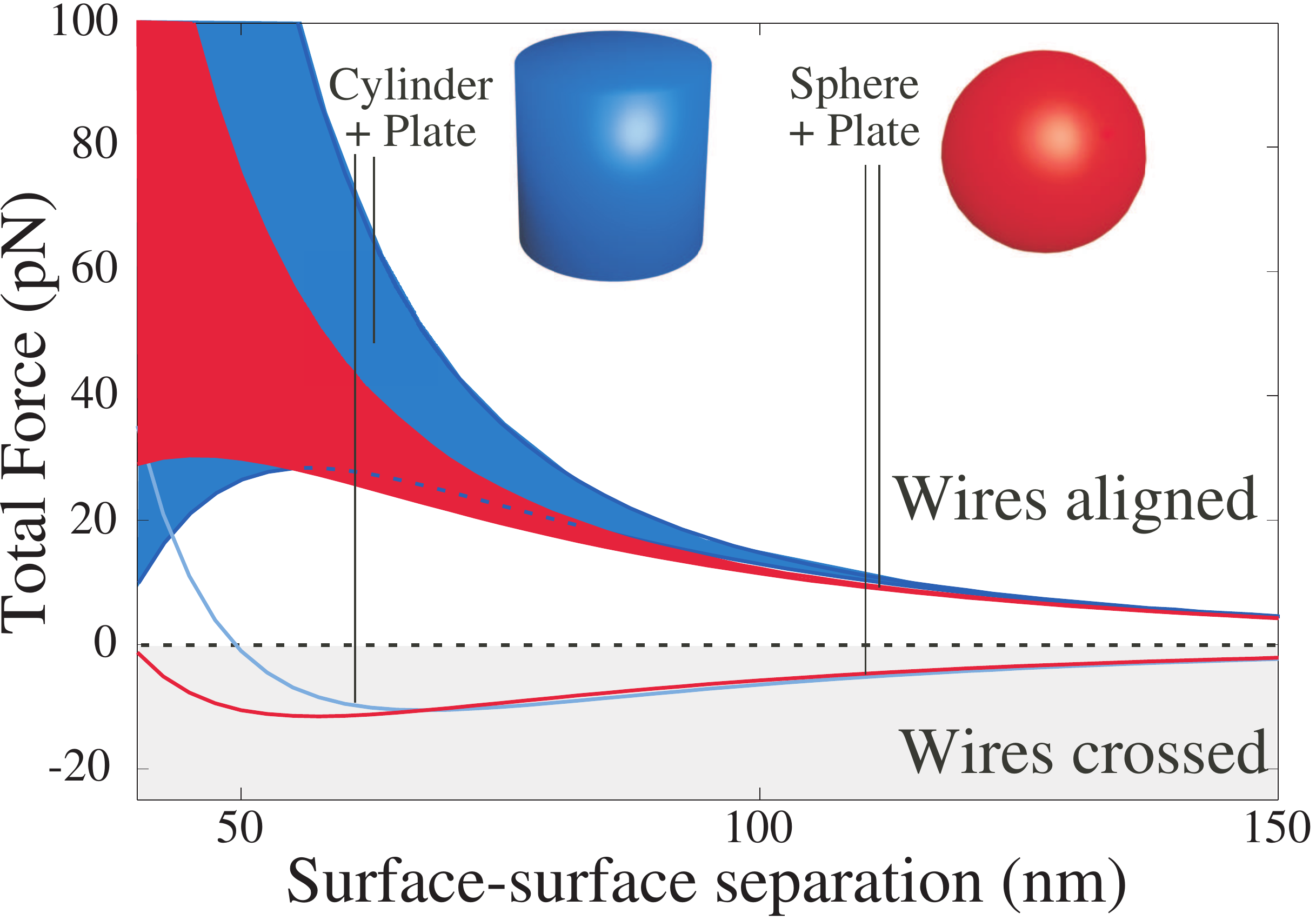}
\centering
\caption{Forces for sphere-plate (red) and cylinder-plate
  configurations, for $a = 100\,\mathrm{nm}$.  In both cases, the wire
  pattern of~\figref{config} is stamped onto the surface of both
  objects, which have radii of curvature $R = 200\,\mu\mathrm{m}$.
  The cylinder length is chosen to be $L = 16\,\mu\mathrm{m}$ so that
  the two force curves have comparable magnitude; however the
  cylinder-plate force is proportional to $L$ and so will be much
  larger in magnitude for a more realistic $L \sim R$.}
\label{fig:sphere-cylinder-PFA}
\end{figure}

Results are shown in~\figref{sphere-cylinder-PFA} for $R =
200\,\mu\mathrm{m}$, and are plotted in units of pico-Newtons.  The
sphere-plate force has a peak repulsion of $\sim 10\,\mathrm{pN}$,
which although below the detection limit of current measurement
techniques in fluids, may be observable in the near future.  We also
note that, although the presence of Casimir repulsion may be
experimentally challenging,~\eqref{Intro_bound} can be verified with
much less sensitivity (e.g., $\sim 50\,\mathrm{pN}$ at $d =
50\,\mathrm{nm}$).  For the cylinder-plate, we take a length $L =
16\,\mu\mathrm{m}$ in order for the force to be comparable in
magnitude to the sphere-plate force.  However, it is clear that if we
instead take $L \sim R$, we can obtain both repulsive and attractive
forces on the order of $10^3\,\mathrm{pN}$, well within current
experimental detection ranges.  Therefore, this system seems most
attractive for detection of an attractive-repulsive transition with
orientation in a fluid if techniques similar to those
of~\citeasnoun{Wei10} to align the long axis of the cylinder relative
to the plane can be extended to fluid-separated objects.

\section{Conclusions}

We have presented an example system, consisting of microstructured
slabs, in which PFA fails at relatively small separations.  Instead,
EMA qualitatively describes the behavior of the Casimir force,
including the case in which the force can be switched from attractive
to repulsive as the slabs are rotated.  We have discussed the
prospects for detecting these forces in experiments.  One issue
arising in an experiment, not discussed here, is the increased
complexity of the electrostatic calibrations.  Conceivably, these
complications can be eliminated in the fluid-separated case by the
addition of electrolytes to the fluid.  This should also reduce or
eliminate the effect of accumulated surface charges on the gold-silica
interface, due to contact potentials~\cite{Man10, Man10:PRA}.  For the
case of vacuum-separated doped silicon, a more involved calibration
procedure is required.

Although not examined here, we also note that orientational
attractive-repulsive transitions may also occur using naturally
anisotropic materials.  We have confirmed this for the case of lithium
niobate slabs immersed in ethanol at zero temperature; although here
$d_c > 7\,\mu\mathrm{m}$ (implying that finite-temperature effects
must be taken into account), transitions at smaller separations may be
possible with appropriate materials.

\section{Acknowledgments}

We thank D. Woolf, J. Munday, F. Intravaia, and M. Maghrebi for useful
discussions, and R. Zhao and K. Berggren for helpful references.  This
work was supported by the Army Research Office through the ISN under
Contract No. W911NF-07-D-0004, and by DARPA under contract
N66001-09-1-2070-DOD and under DOE/NNSA contract DE-AC52-06NA25396.

\appendix

\section{Details of Computational Methods}
\label{sec:appA}

\subsection{Scattering from periodic arrays}
\label{sec:Method-scattering}

There are numerous computational and semi-analytical methods from
classical electromagnetic scattering that can be adapted to Casimir
calculations~\cite{Steven11:review}.  For example, there are
computational techniques based on generic grids/meshes,
e.g. finite-difference methods~\cite{Rodriguez07:PRA, Pasquali09,
  RodriguezMc09:PRA, McCauleyRo10:PRA}.  There are also Casimir
methods~\cite{Lambrecht09,Davids10} based on classical cross-section methods
(rigorous coupled-wave analysis~\cite{Moharam95}, also called
eigenmode expansion~\cite{Bienstman01}), which divide the geometry
into slices with constant cross-sections and match modal expansions
between slices.  Alternatively, there are spectral integral-equation
methods tracing their roots to classic Mie scattering and related
problems~\cite{Stratton41}: one divides the geometry into
high-symmetry objects like spheres and cylinders, computing the
scattering matrix for each object in a specialized basis
(e.g. spherical waves), and then combining the matrices from different
objects to match the boundary conditions. These methods have been
adapted to Casimir problems for geometries consisting of a finite
number of isolated objects~\cite{Emig07,Neto08,Kenneth08,Rahi09:PRD}
and corrugated surfaces~\cite{emig04_2}.  Here, we adapt similar
methods to periodic arrays of isolated objects (cylinders) by
exploiting classical lattice-sum scattering methods~\cite{Yasumoto04}.
For completeness, and because our formulation is directly in imaginary
frequency and our normalization conventions differ from other
applications~\cite{Kushta00, Yasumoto04}, we outline the derivation in
this section.

The scattering matrix $\mathbb{T}_0$ from a single circular cylinder can
be computed analytically in the basis of cylindrical multipoles, using
the basis functions~\cite{Rahi09:PRD}:

\begin{eqnarray}
|\vec{E}\rangle_{k_z,n}^{p, \mathrm{reg}/\mathrm{out}}(i\xi,\vec{x}) &=& \mathbb{L}^p \left[\phi^{\mathrm{reg}/ \mathrm{out}}_{k_z,n}\left(i\xi,\vec{x}-\vec{x}_j\right)\hat{\vec{z}}\right] \nonumber \\
\label{eq:basis}
\end{eqnarray}
where the linear differential operators $\mathbb{L}^p$ ($p = M,N$ are
the transverse electric and transverse magnetic polarizations,
respectively) are:

\begin{eqnarray*}
\mathbb{L}^M &=& \frac{1}{\sqrt{k_z^2 + \xi^2}} \vec{\nabla} \times\\
\mathbb{L}^N &=& \frac{1}{\xi\sqrt{k_z^2+\xi^2}} \vec{\nabla} \times \vec{\nabla}\times
\end{eqnarray*}

and the cylindrical wave functions are:

\begin{eqnarray*}
\phi^{\mathrm{reg}}_{k_z,n}\left(i\xi,\vec{x}\right) &=& I_n\left(\sqrt{k_z^2+\xi^2}\sqrt{x^2+y^2}\right)e^{ik_zz+in\theta} \\
\phi^{\mathrm{out}}_{k_z,n}\left(i\xi,\vec{x}\right) &=& K_n\left(\sqrt{k_z^2+\xi^2}\sqrt{x^2+y^2}\right)e^{ik_zz+in\theta}
\end{eqnarray*}

$I_n$ and $K_n$ are the modified cylindrical Bessel functions, and
$\vec{x}_j$ gives the coordinate origin.  The calculation of
$\mathbb{T}_0$ is straightforward and we omit it here.  We note that
although the present method is simplified by the semi-analytic
calculation of the scattering from circular cylinders, methods exist
that can compute the scattering matrices of non-circular cylinders of
arbitrary cross-section~\cite{Reid11:2D}; such a hybrid method allows
for efficient scattering computations from general two-dimensional
arrays of compact objects.

We now derive the scattering matrix $\mathbb{T}$ for a periodic array
of cylindrical scatters in a uniform medium (the fluid interface will
be added later).  $\mathbb{T}$ is defined to be the scattering matrix
from each cylinder $j$ in the basis given by~\eqref{basis} with origin
$\vec{x}_j = j a \hat{\vec{y}}$, in the presence of the entire array
of cylinders.  This array has period $a$ in the $y$-direction and is
translation-invariant in $z$, and only a single layer in the
$x$-direction.  In other words, for an incident field
$|\vec{E}\rangle^\mathrm{in}$, the scattered field emitted by
currents on cylinder $j$ is:

\begin{equation}
|\vec{E}^{(j)}\rangle^\mathrm{scatt} =
e^{i k_y j a} \mathbb{T}|\vec{E}\rangle^\mathrm{in}
\end{equation}

Here we have assumed that the incident field
$|\vec{E}\rangle^\mathrm{in}$ is a planewave with transverse
wavevector components $k_y$ and $k_z$, and have normalized it with
respect to the origin.  Then using the translation
matrices~\cite{Rahi09:PRD} for the basis functions of~\eqref{basis},
the scattered field from object $j$, converted to the basis at the
origin ($j = 0$), is given by:

\begin{equation}
^{p,\mathrm{reg}}_{k_z, n}\langle \vec{E}^{(j=0)} |\vec{E}^{(j)}\rangle
=
\sum_{n^{\prime\prime}, n^\prime}
\vec{S}^{(j)}_{n - n^\prime} e^{i k_y j a} (-1)^{n^\prime}
\mathbb{T}^{p,p^\prime}_{n^\prime,n^{\prime\prime}}
 ~^{p^\prime}_{n^{\prime\prime}}\langle \vec{E} | \vec{E} \rangle^\mathrm{in}
\end{equation}
where the elements of the lattice sum $\vec{S}$ are given by:
\begin{equation}
S^{(j)}_{m} = K_m\left(\left| \vec{x}\right|\sqrt{k_z^2 + \varepsilon_2 \xi^2}\right) e^{-im \theta_j}
\end{equation}
and $\theta_j = \mathrm{sign}(j)\frac{\pi}{2}$; and this
transformation is diagonal in polarization $p$.  Using linearity of
the scattering process to sum over all incident cylindrical waves, the
total incoming field at the origin $x_0$ for a planewave
$|\vec{E}\rangle^\mathrm{in}$ is obtained by summing the incident
fields and the scattered fields from all cells $j\neq 0$:

\begin{eqnarray*}
|\vec{E}\rangle^{\mathrm{in, total}} & = & \left[1 + \left(\sum_{j\neq 0} e^{i k_y j a}\mathbb{S}^{(j)} \right) \mathbb{A} \mathbb{T} \right] |\vec{E}\rangle^{\mathrm{in}}\\
& = & \left(1 + \mathbb{S}\mathbb{A}\mathbb{T}\right) |\vec{E}\rangle^{\mathrm{in}}
\label{eq:lattice_sum}
\end{eqnarray*}
The matrix $\mathbb{S}$ is formally defined as the infinite sum of all
$\mathbb{S}^{(j)}$ (in imaginary frequency this sum is well-defined
due to the exponential decay of $K_n$), and $\mathbb{A}_{n,n^\prime} =
(-1)^{n}\delta_{n,n^\prime}$.  The field scattered from the cylinder
at $\vec{x} = \vec{x}_0$, in the presence of the array of cylinders,
is then:

\begin{eqnarray*}
|\vec{E}^{(j=0)}\rangle^{\mathrm{scatt}} &=& \mathbb{T}_0 \left(1 + \mathbb{S}\mathbb{A}\mathbb{T}\right) |\vec{E}\rangle^\mathrm{in}\\
&=& \mathbb{T} |\vec{E}\rangle^\mathrm{in}
\end{eqnarray*}
where the last equality follows by the definition of $\mathbb{T}$.
Because this equation holds for arbitrary incident field vectors
$|\vec{E}\rangle^\mathrm{in}$, the equation holds for the operators as
well, allowing us to solve for $\mathbb{T}$:

\begin{equation}
\mathbb{T} = \left[ 1 - \mathbb{T}_0 \mathbb{S} \mathbb{A}\right]^{-1} \mathbb{T}_0
\label{eq:full_Tmat}
\end{equation}
Although both $\mathbb{S}$ and $\mathbb{A}$ are diagonal in
polarization $p$, in general $\mathbb{T}_0$ is not and different
polarizations will couple to the periodicity via~\eqref{full_Tmat}.
The scattering matrix from the entire array is a sum of $\mathbb{T}$
over all unit cells.  This sum is more naturally expressed in a
planewave basis, using the wave conversion matrices $\mathbb{D}$ for
cylindrical waves~\cite{Rahi09:PRD} (here we have absorbed the
normalization factors $C_p$, defined in Appendix
B. of~\citeasnoun{Rahi09:PRD}, into the definition of $\mathbb{D}$),
from which we find the planewave scattering matrix:

\begin{equation}
\mathbb{R}_\vec{k} = \mathbb{D}_\vec{k}^\dagger \mathbb{T} \mathbb{D}_\vec{K}
\label{eq:planar_Tmat}
\end{equation}
Here $\vec{k}$ is the reduced Bloch vector, and the matrix
$\mathbb{R}_\vec{k}$ couples all vectors $\vec{k}+m \pi \hat{\vec{y}}$
for all integer $m$. 

The incorporation of multiple interfaces in the $x$-direction is
achieved via a standard transfer matrix approach, e.g.,
\citeasnoun{Yasumoto04}.  For the case of a uniform medium
$\varepsilon_3$ outside of the plates, the Fresnel formula are
combined in a straightforward manner with the matrix $\mathbb{T}$
above to finally give the full scattering matrix $\mathbb{R}$ for the
objects of~\figref{config}.

For numerical computations, ten cylindrical multipole moments in the
single-cylinder scattering matrix, $10/\xi$ terms in the lattice sum
of~\eqref{lattice_sum}, and a simple exponential extrapolation from
only $7\times 7$ Brillouin zones for~\eqref{planar_Tmat} were
sufficient for $< 1\%$ error.

\subsection{Effective-Medium Approximation}
\label{sec:Method-EMA}

In this section we outline the method used to predict the scattering
matrices $\mathbb{R}$ from a specified $\stackrel{\leftrightarrow}{\varepsilon}$.
The method is given in full detail in~\citeasnoun{Rosa08}.

In an anisotropic EMA, we assume that the i-th object is described by
a homogeneous (position-independent) permittivity tensor
\begin{eqnarray}
\label{eq:anisotropy}
\stackrel{\leftrightarrow}{\varepsilon} = \left[ 
\begin{array}{ccc}
\varepsilon_{xx} & \varepsilon_{xy} & \varepsilon_{xz} \\
\varepsilon_{xy} & \varepsilon_{yy} & \varepsilon_{yz} \\
\varepsilon_{xz} & \varepsilon_{yz} & \varepsilon_{zz}
\end{array} \right], 
\end{eqnarray}
where the frequency dependence is left implicit and we used the fact
that the permittivity tensor is symmetric on its spatial
indices~\cite{Landau:EM}. The relatively symmetric character of our
setup leads to a very useful simplification in~\eqref{anisotropy} that
becomes apparent in a coordinate system with its x-axis perpendicular
to the slabs interfaces: it is clear that regardless of the in-plane
orientation of the wires there is no mixing with the $x$-component,
and so
\begin{equation}
\varepsilon_{xy} =  \varepsilon_{xz} = 0 .
\end{equation}
In addition, for our simplified EMA we make the further approximation
that $\varepsilon_{xx} = \varepsilon_{yy} \equiv \varepsilon_\perp $.
This symmetry implies that the $\varepsilon$ tensor is diagonal in the
$x,y,z$ coordinate system of~\figref{config} (top) and is of the form:
\begin{eqnarray}
\label{eq:anisotropyDiagonal}
\stackrel{\leftrightarrow}{\varepsilon} = \left[ 
\begin{array}{ccc}
\varepsilon_\perp & 0 & 0 \\
0 & \varepsilon_\perp  & 0 \\
0 & 0 & \varepsilon_{||}
\end{array} \right] 
\end{eqnarray}
Consider a planewave of transverse wavevector $\vec{k}_\perp = (k_y,
k_z)$ impinging on the slab, and let $\phi$ be the angle
$\vec{k}_\perp$ makes with the $z$-axis.  For our purpose, the
scattering problem is best solved in the coordinate system $(x^\prime,
y^\prime, z^\prime)$ of the plane wave, where $x^\prime = x$ and
$z^\prime$ is parallel to $\vec{k}_\perp$.  In this basis, the
$\stackrel{\leftrightarrow}{\varepsilon}$ tensor becomes:
\begin{eqnarray*}
\label{eq:anisotropy2}
\stackrel{\leftrightarrow}{\varepsilon} =
\left[ 
\begin{array}{ccc}
\varepsilon_\perp & 0                     & 0 \\
0        & \varepsilon_\perp c^2  +  \varepsilon_{||}s^2                & (\varepsilon_{||}-\varepsilon_\perp) s c \\
0                                      & \left(\varepsilon_{||}-\varepsilon_\perp\right) s c & \varepsilon_{||} c^2 + \varepsilon_\perp s^2
\end{array} \right]
\end{eqnarray*}
where $c = \cos\phi$ and $s = \sin\phi$.

We can now proceed to determine the scattering matrices of planewaves
for the isotropic-anisotropic flat interface.  After a considerably
long algebraic calculation (for details see~\citeasnoun{Rosa08}, but
note the change in coordinate systems between our~\figref{config} and
their Fig. 3, and that for us $c = \cos \phi$ and the speed of light
is set to 1), it is possible to show that the four reflected/incident
amplitude ratios are given by
\begin{eqnarray*}
r^{\rm TE,TE}(i \xi, {\bf k}_{\perp})  & = & \left. \frac{\det \mathbb{M}_1}{\det \mathbb{M}}  \right|_{\stackrel{\omega \rightarrow i \xi}{k_{z'}\rightarrow k_{\perp}}} \\
r^{\rm TM,TE} (i \xi, {\bf k}_{\perp}) & = & \left. \frac{\det \mathbb{M}_2}{\det \mathbb{M}}  \right|_{\stackrel{\omega \rightarrow i \xi}{k_{z'}\rightarrow k_{\perp}}}  \\
r^{\rm TE,TM} (i \xi, {\bf k}_{\perp}) & = & \left. \frac{\det \mathbb{M}_3}{\det \mathbb{M}}  \right|_{\stackrel{\omega \rightarrow i \xi}{k_{z'}\rightarrow k_{\perp}}} \\
r^{\rm TM,TM} (i \xi, {\bf k}_{\perp}) & = & \left. \frac{\det \mathbb{M}_4}{\det \mathbb{M}}  \right|_{\stackrel{\omega \rightarrow i \xi}{k_{z'}\rightarrow k_{\perp}}}
\end{eqnarray*}
where
\begin{eqnarray*}
&&\mathbb{M} = \left[ \begin{array}{cccc}
  -1 & 0 & \alpha^{(1)} & \alpha^{(2)} \\    
    q_{\rm in}/\omega & 0 & -\beta^{(1)} & -\beta^{(2)} \\
  0 &   q_{\rm in}/\omega & 1 & 1 \\
  0 & -1 & \gamma^{(1)} & \gamma^{(2)}  
\end{array} \right] \nonumber \\
\end{eqnarray*}

\begin{eqnarray*}
&&\mathbb{M}_1 = \left[ \begin{array}{cccc}
  1 & 0 & \alpha^{(1)} & \alpha^{(2)} \\    
    q_{\rm in}/\omega & 0 & -\beta^{(1)} & -\beta^{(2)} \\
  0 &   q_{\rm in}/\omega  & 1 & 1 \\
  0 & -1 & \gamma^{(1)} & \gamma^{(2)}  
\end{array} \right]  \nonumber \\
\end{eqnarray*}

\begin{eqnarray*}
&&\mathbb{M}_2 = \left[ \begin{array}{cccc}
  -1 & 1 & \alpha^{(1)} & \alpha^{(2)} \\    
    q_{\rm in}/\omega  &   q_{\rm in}/\omega  & -\beta^{(1)} & -\beta^{(2)} \\
  0 & 0 & 1 & 1 \\
  0 & 0 & \gamma^{(1)} & \gamma^{(2)}  
\end{array} \right]  \nonumber \\
\end{eqnarray*}

\begin{eqnarray*}
&&\mathbb{M}_3 = \left[ \begin{array}{cccc}
  0 & 0 & \alpha^{(1)} & \alpha^{(2)} \\    
  0 & 0 & -\beta^{(1)} & -\beta^{(2)} \\
    q_{\rm in}/\omega  &   q_{\rm in}/\omega  & 1 & 1 \\
  1 & -1 & \gamma^{(1)} & \gamma^{(2)}  
\end{array} \right] \nonumber \\
\end{eqnarray*}

\begin{eqnarray*}
&&\mathbb{M}_4 = \left[ \begin{array}{cccc}
  -1 & 0 & \alpha^{(1)} & \alpha^{(2)} \\    
    q_{\rm in}/\omega  & 0 & -\beta^{(1)} & -\beta^{(2)} \\
  0 &   q_{\rm in}/\omega  & 1 & 1 \\
  0 & 1 & \gamma^{(1)} & \gamma^{(2)}  
\end{array} \right] 
\end{eqnarray*}
Here $q_\mathrm{in} \equiv \sqrt{\omega^2 - \vec{k}_\perp^2}$; the
coefficients in the matrices above are defined by
\begin{eqnarray}
&&\alpha^{(p)} = \frac{(q^{(p)})^2 - \omega^2 A}{\omega^2 C_1}  \nonumber \\
&&\beta^{(p)} =  -\omega \frac{L_{31}}{q^{(p)}} - \omega \frac{L_{32}}{q^{(p)}} \alpha^{(p)} \nonumber \\
&&\gamma^{(p)} = -\omega \frac{L_{41}}{q^{(p)}} - \omega \frac{L_{42}}{q^{(p)}} \alpha^{(p)} 
\label{eq:greeks} 
\end{eqnarray} 
where $p=1,2$ and also
\begin{equation}
\label{eq:bi-solutions}
q^{(p)} = q^{(\pm)} = \omega \frac{1}{\sqrt{2}} \sqrt{ A+B \pm \sqrt{(A-B)^2 + 4 C}} 
\end{equation}
with
\begin{eqnarray*}
A &=&  L_{14} L_{41}  \nonumber \\
B &=& L_{23} L_{32}  \nonumber \\
C &=&  L_{14} L_{42} L_{23} L_{31} 
\end{eqnarray*}
Finally, the elements $L_{ij}$, are defined in terms of the
permittivity tensor as
\begin{eqnarray*}
&&L_{14} =  \frac{k_{z'}^2 }{ \omega^2 \varepsilon_\perp} - 1  \nonumber \\
&&L_{23}= 1 \nonumber \\
&&L_{31}=-L_{42}= (\varepsilon_\perp - \varepsilon_{||}) s c   \nonumber \\
&&L_{32}=- \frac{k_{z'}^2 }{\omega^2} + \varepsilon_\perp c^2 + \varepsilon_{||} s^2 \nonumber \\
&&L_{41}= -\varepsilon_{||} c^2 - \varepsilon_\perp s^2
\end{eqnarray*}

A careful analysis shows that the equations~\eqref{greeks} become
singular at $\phi = n \frac{\pi}{2}$ for integer $n$.  In this case,
we take the limit $\phi \rightarrow n\frac{\pi}{2}$ numerically.

\section{Suppression of non-specular scattering}
\label{sec:appB}
\begin{figure}[tb]
\includegraphics[width=1.0\columnwidth]{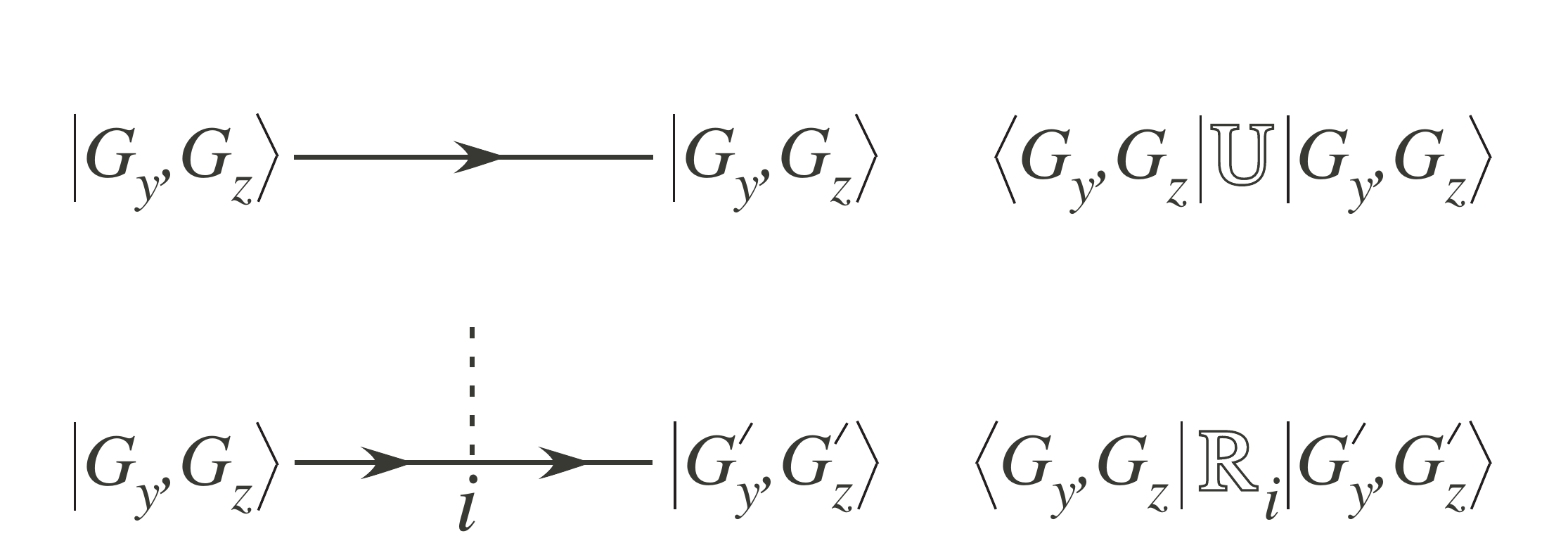}
\centering
\caption{Feynman rules for the diagrammatic expansion of the Casimir
  energy, following~\citeasnoun{Maghrebi10:arxiv}.  In our notation,
  planewave states are indexed by their reciprocal lattice vector
  $\vec{G}_\perp = \left( G_y, G_z\right)$; frequency, polarization,
  and the conserved reduced Bloch vector $\vec{k}_\perp$ are
  suppressed as our focus is on non-specular (i.e., $\vec{G}_\perp$
  not conserved) reflections.  Here $\mathbb{U}$ is the free-space
  translation matrix, and $\mathbb{R}_i$ the scattering matrix for
  object $i$ (= 1, 2 for our case).  Primes on the polarizations
  (e.g., $y^\prime$) denote distinct wavevectors.}
\label{fig:Feynman-rules}
\end{figure}

\begin{figure}[tb]
\includegraphics[width=1.0\columnwidth]{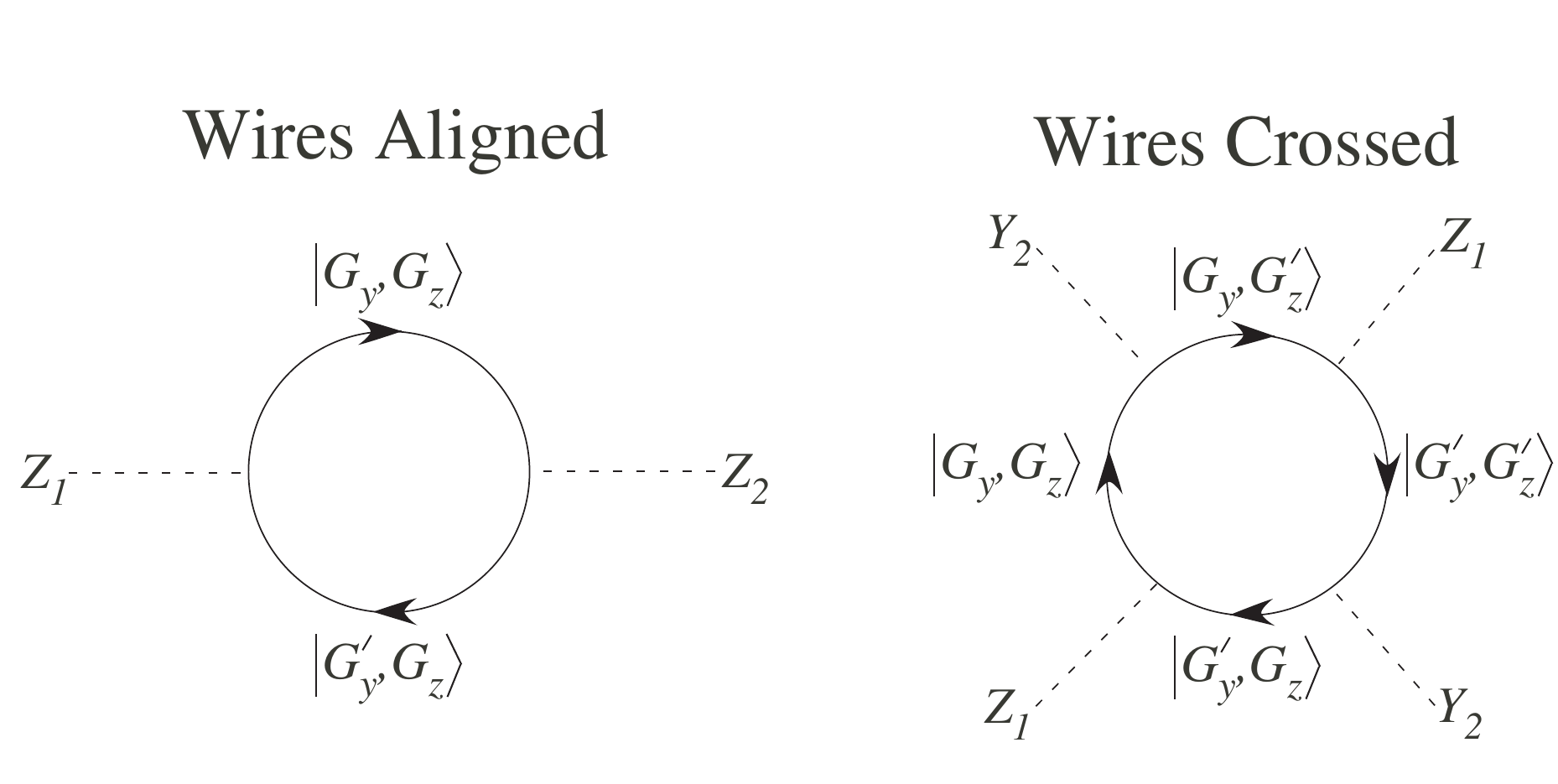}
\centering
\caption{Lowest-order diagrams representing the contribution of
  non-specular scattering events to the Casimir energy, constructed
  using the rules of~\figref{Feynman-rules}.  Because we are measuring
  the energy of interaction between two objects, successive
  interaction vertices must involve different objects.  On the left,
  when both objects 1 and 2 are directed along the $z$-axis, a
  non-specular event can occur in a 2-vertex diagram.  However, when
  the wires are crossed, the lowest-order allowed non-specular process
  involves a 4-vertex diagram.}
\label{fig:scattering}
\end{figure}

In this section we give a proof that non-specular scatterings (i.e.,
those not conserving the perpendicular wavevector $\vec{k}_\perp +
\vec{G}_\perp$) are greatly suppressed for crossed slabs relative to
aligned slabs.  This explains the wide range of validity for the
specular-reflection approximation for crossed slabs.  The proof is a
straightforward application of a recently-developed diagrammatic
expansion for the Casimir force~\cite{Maghrebi10:arxiv}.  In this
framework, the $\log \det$ expression of~\eqref{energy} is
re-expressed as $\mathrm{Tr} \log$ and expanded in a power series.
Each term in the power series can be computed via a set of Feynman
rules, shown in~\figref{Feynman-rules}.  For sufficiently large
separations, this series is rapidly convergent.

Each scattering event is represented as the process $|G_x, G_z\rangle
\rightarrow |G_x^\prime, G_z^\prime\rangle$, where $G_x, G_z$
represent the perpendicular wavevector of the planewave.  We suppress
both the frequency and reduced Bloch vector $\vec{k}_\perp$ (which are
both conserved in the scattering process), and the polarization $p$
(which is not relevant to our analysis of non-specular reflection).
Each object is represented as a pair indicating the object alignment
$(Y, Z)$ and object index $(1, 2)$.  The key point is that, due to
translation-invariance, the scattering matrix for a $Z$ object is
proportional to $\delta\left(G_z - G_z^\prime\right)$, and for a $Y$
object is proportional to $\delta\left(G_y - G_y^\prime\right)$.
Further, the free-space propagator is diagonal in the planewave basis
and thus conserves both $G_y$ and $G_z$.  Successive interaction
vertices in a diagram must involve distinct objects as external
insertions; therefore, for the crossed configuration, a planewave must
scatter twice as many times to return to its original state in the
lowest-order diagram relative to the aligned configuration.  Example
lowest-order diagrams are shown in~\figref{scattering}.  We find that
for aligned (e.g., $ZZ$) slabs, the lowest order diagram involving
non-specular reflection involves two vertices, while for crossed slabs
it involves four vertices.  Since each propagator represents an
exponential attenuation, and each scattering event multiplication by a
number of magnitude less than unity, this implies that non-specular
interactions are greatly suppressed for crossed slabs.


\begin{thebibliography}{59}
\expandafter\ifx\csname natexlab\endcsname\relax\def\natexlab#1{#1}\fi
\expandafter\ifx\csname bibnamefont\endcsname\relax
  \def\bibnamefont#1{#1}\fi
\expandafter\ifx\csname bibfnamefont\endcsname\relax
  \def\bibfnamefont#1{#1}\fi
\expandafter\ifx\csname citenamefont\endcsname\relax
  \def\citenamefont#1{#1}\fi
\expandafter\ifx\csname url\endcsname\relax
  \def\url#1{\texttt{#1}}\fi
\expandafter\ifx\csname urlprefix\endcsname\relax\def\urlprefix{URL }\fi
\providecommand{\bibinfo}[2]{#2}
\providecommand{\eprint}[2][]{\url{#2}}

\bibitem[{\citenamefont{Hertzberg et~al.}(2005)\citenamefont{Hertzberg, Jaffe,
  Kardar, and Scardicchio}}]{Hertzberg05}
\bibinfo{author}{\bibfnamefont{M.~P.} \bibnamefont{Hertzberg}},
  \bibinfo{author}{\bibfnamefont{R.~L.} \bibnamefont{Jaffe}},
  \bibinfo{author}{\bibfnamefont{M.}~\bibnamefont{Kardar}}, \bibnamefont{and}
  \bibinfo{author}{\bibfnamefont{A.}~\bibnamefont{Scardicchio}},
  \bibinfo{journal}{Phys. Rev. Lett.} \textbf{\bibinfo{volume}{95}},
  \bibinfo{pages}{250402} (\bibinfo{year}{2005}).

\bibitem[{\citenamefont{Rodrigues et~al.}(2006)\citenamefont{Rodrigues,
  Maia~Neto, Lambrecht, and Reynaud}}]{Rodrigues06:torque}
\bibinfo{author}{\bibfnamefont{R.~B.} \bibnamefont{Rodrigues}},
  \bibinfo{author}{\bibfnamefont{P.~A.} \bibnamefont{Maia~Neto}},
  \bibinfo{author}{\bibfnamefont{A.}~\bibnamefont{Lambrecht}},
  \bibnamefont{and} \bibinfo{author}{\bibfnamefont{S.}~\bibnamefont{Reynaud}},
  \bibinfo{journal}{Europhys. Lett.} \textbf{\bibinfo{volume}{75}},
  \bibinfo{pages}{822} (\bibinfo{year}{2006}).

\bibitem[{\citenamefont{Rodriguez
  et~al.}(2007{\natexlab{a}})\citenamefont{Rodriguez, Ibanescu, Iannuzzi,
  Capasso, Joannopoulos, and Johnson}}]{Rodriguez07:PRL}
\bibinfo{author}{\bibfnamefont{A.}~\bibnamefont{Rodriguez}},
  \bibinfo{author}{\bibfnamefont{M.}~\bibnamefont{Ibanescu}},
  \bibinfo{author}{\bibfnamefont{D.}~\bibnamefont{Iannuzzi}},
  \bibinfo{author}{\bibfnamefont{F.}~\bibnamefont{Capasso}},
  \bibinfo{author}{\bibfnamefont{J.~D.} \bibnamefont{Joannopoulos}},
  \bibnamefont{and} \bibinfo{author}{\bibfnamefont{S.~G.}
  \bibnamefont{Johnson}}, \bibinfo{journal}{Phys. Rev. Lett.}
  \textbf{\bibinfo{volume}{99}}, \bibinfo{pages}{080401}
  (\bibinfo{year}{2007}{\natexlab{a}}).

\bibitem[{\citenamefont{Emig}(2007)}]{Emig07:ratchet}
\bibinfo{author}{\bibfnamefont{T.}~\bibnamefont{Emig}}, \bibinfo{journal}{Phys.
  Rev. Lett.} \textbf{\bibinfo{volume}{98}}, \bibinfo{pages}{160801}
  (\bibinfo{year}{2007}).

\bibitem[{\citenamefont{Rodriguez et~al.}(2008)\citenamefont{Rodriguez,
  Joannopoulos, and Johnson}}]{RodriguezJo08:PRA}
\bibinfo{author}{\bibfnamefont{A.~W.} \bibnamefont{Rodriguez}},
  \bibinfo{author}{\bibfnamefont{J.~D.} \bibnamefont{Joannopoulos}},
  \bibnamefont{and} \bibinfo{author}{\bibfnamefont{S.~G.}
  \bibnamefont{Johnson}}, \bibinfo{journal}{Phys. Rev.~A}
  \textbf{\bibinfo{volume}{77}}, \bibinfo{pages}{062107}
  (\bibinfo{year}{2008}).

\bibitem[{\citenamefont{Rosa et~al.}(2008{\natexlab{a}})\citenamefont{Rosa,
  Dalvit, and Milonni}}]{Rosa08:PRL}
\bibinfo{author}{\bibfnamefont{F.~S.~S.} \bibnamefont{Rosa}},
  \bibinfo{author}{\bibfnamefont{D.~A.~R.} \bibnamefont{Dalvit}},
  \bibnamefont{and} \bibinfo{author}{\bibfnamefont{P.~W.}
  \bibnamefont{Milonni}}, \bibinfo{journal}{Phys. Rev. Lett.}
  \textbf{\bibinfo{volume}{100}}, \bibinfo{pages}{183602}
  (\bibinfo{year}{2008}{\natexlab{a}}).

\bibitem[{\citenamefont{Rahi and Zaheer}(2010)}]{RahiZa10}
\bibinfo{author}{\bibfnamefont{S.~J.} \bibnamefont{Rahi}} \bibnamefont{and}
  \bibinfo{author}{\bibfnamefont{S.}~\bibnamefont{Zaheer}},
  \bibinfo{journal}{Phys. Rev. Lett.} \textbf{\bibinfo{volume}{104}},
  \bibinfo{pages}{070405} (\bibinfo{year}{2010}).

\bibitem[{\citenamefont{Levin et~al.}(2010)\citenamefont{Levin, McCauley,
  Rodriguez, Reid, and Johnson}}]{LevinMc10}
\bibinfo{author}{\bibfnamefont{M.}~\bibnamefont{Levin}},
  \bibinfo{author}{\bibfnamefont{A.~P.} \bibnamefont{McCauley}},
  \bibinfo{author}{\bibfnamefont{A.~W.} \bibnamefont{Rodriguez}},
  \bibinfo{author}{\bibfnamefont{M.~T.~H.} \bibnamefont{Reid}},
  \bibnamefont{and} \bibinfo{author}{\bibfnamefont{S.~G.}
  \bibnamefont{Johnson}}, \bibinfo{journal}{Phys. Rev. Lett.}
  \textbf{\bibinfo{volume}{105}}, \bibinfo{pages}{090403}
  (\bibinfo{year}{2010}).

\bibitem[{\citenamefont{de~Man et~al.}(2009)\citenamefont{de~Man, Heeck,
  Wijngaarden, and Iannuzzi}}]{Man09}
\bibinfo{author}{\bibfnamefont{S.}~\bibnamefont{de~Man}},
  \bibinfo{author}{\bibfnamefont{K.}~\bibnamefont{Heeck}},
  \bibinfo{author}{\bibfnamefont{R.~J.} \bibnamefont{Wijngaarden}},
  \bibnamefont{and} \bibinfo{author}{\bibfnamefont{D.}~\bibnamefont{Iannuzzi}},
  \bibinfo{journal}{Phys. Rev. Lett.} \textbf{\bibinfo{volume}{103}},
  \bibinfo{pages}{040402} (\bibinfo{year}{2009}).

\bibitem[{\citenamefont{Feiler et~al.}(2008)\citenamefont{Feiler, Bergstrom,
  and Rutland}}]{Feiler08}
\bibinfo{author}{\bibfnamefont{A.~A.} \bibnamefont{Feiler}},
  \bibinfo{author}{\bibfnamefont{L.}~\bibnamefont{Bergstrom}},
  \bibnamefont{and} \bibinfo{author}{\bibfnamefont{M.~W.}
  \bibnamefont{Rutland}}, \bibinfo{journal}{Langmuir}
  \textbf{\bibinfo{volume}{24}}, \bibinfo{pages}{2274} (\bibinfo{year}{2008}).

\bibitem[{\citenamefont{Munday et~al.}(2009)\citenamefont{Munday, Capasso, and
  Parsegian}}]{Munday09}
\bibinfo{author}{\bibfnamefont{J.}~\bibnamefont{Munday}},
  \bibinfo{author}{\bibfnamefont{F.}~\bibnamefont{Capasso}}, \bibnamefont{and}
  \bibinfo{author}{\bibfnamefont{V.~A.} \bibnamefont{Parsegian}},
  \bibinfo{journal}{Nature} \textbf{\bibinfo{volume}{457}},
  \bibinfo{pages}{170} (\bibinfo{year}{2009}).

\bibitem[{\citenamefont{Rodriguez et~al.}(2010)\citenamefont{Rodriguez,
  McCauley, Woolf, Capasso, Joannopoulos, and Johnson}}]{RodriguezMc10:PRL}
\bibinfo{author}{\bibfnamefont{A.~W.} \bibnamefont{Rodriguez}},
  \bibinfo{author}{\bibfnamefont{A.~P.} \bibnamefont{McCauley}},
  \bibinfo{author}{\bibfnamefont{D.}~\bibnamefont{Woolf}},
  \bibinfo{author}{\bibfnamefont{F.}~\bibnamefont{Capasso}},
  \bibinfo{author}{\bibfnamefont{J.~D.} \bibnamefont{Joannopoulos}},
  \bibnamefont{and} \bibinfo{author}{\bibfnamefont{S.~G.}
  \bibnamefont{Johnson}}, \bibinfo{journal}{Phys. Rev. Lett.}
  \textbf{\bibinfo{volume}{104}}, \bibinfo{pages}{160402}
  (\bibinfo{year}{2010}).

\bibitem[{\citenamefont{Derjaguin et~al.}(1956)\citenamefont{Derjaguin,
  Abrikosova, and Lifshitz}}]{Derjaguin56}
\bibinfo{author}{\bibfnamefont{B.~V.} \bibnamefont{Derjaguin}},
  \bibinfo{author}{\bibfnamefont{I.~I.} \bibnamefont{Abrikosova}},
  \bibnamefont{and} \bibinfo{author}{\bibfnamefont{E.~M.}
  \bibnamefont{Lifshitz}}, \bibinfo{journal}{Q. Rev. Chem. Soc.}
  \textbf{\bibinfo{volume}{10}}, \bibinfo{pages}{295} (\bibinfo{year}{1956}).

\bibitem[{\citenamefont{Kushta and Yasumoto}(2000)}]{Kushta00}
\bibinfo{author}{\bibfnamefont{T.}~\bibnamefont{Kushta}} \bibnamefont{and}
  \bibinfo{author}{\bibfnamefont{K.}~\bibnamefont{Yasumoto}},
  \bibinfo{journal}{PIER} \textbf{\bibinfo{volume}{29}}, \bibinfo{pages}{69}
  (\bibinfo{year}{2000}).

\bibitem[{\citenamefont{Yasumoto and Jia}(2004)}]{Yasumoto04}
\bibinfo{author}{\bibfnamefont{K.}~\bibnamefont{Yasumoto}} \bibnamefont{and}
  \bibinfo{author}{\bibfnamefont{H.}~\bibnamefont{Jia}}, in
  \emph{\bibinfo{booktitle}{Wave Prop., Scatt. and Emission in Complex Media}},
  edited by \bibinfo{editor}{\bibfnamefont{Y.}~\bibnamefont{Jin}}
  (\bibinfo{publisher}{World Scientific}, \bibinfo{address}{Singapore},
  \bibinfo{year}{2004}), pp. \bibinfo{pages}{225--249}.

\bibitem[{\citenamefont{Rahi et~al.}(2009)\citenamefont{Rahi, Emig, Graham,
  Jaffe, and Kardar}}]{Rahi09:PRD}
\bibinfo{author}{\bibfnamefont{S.~J.} \bibnamefont{Rahi}},
  \bibinfo{author}{\bibfnamefont{T.}~\bibnamefont{Emig}},
  \bibinfo{author}{\bibfnamefont{N.}~\bibnamefont{Graham}},
  \bibinfo{author}{\bibfnamefont{R.~L.} \bibnamefont{Jaffe}}, \bibnamefont{and}
  \bibinfo{author}{\bibfnamefont{M.}~\bibnamefont{Kardar}},
  \bibinfo{journal}{Phys. Rev.~D} \textbf{\bibinfo{volume}{80}},
  \bibinfo{pages}{085021} (\bibinfo{year}{2009}).

\bibitem[{\citenamefont{Rosa et~al.}(2008{\natexlab{b}})\citenamefont{Rosa,
  Dalvit, and Milonni}}]{Rosa08}
\bibinfo{author}{\bibfnamefont{F.~S.~S.} \bibnamefont{Rosa}},
  \bibinfo{author}{\bibfnamefont{D.~A.~R.} \bibnamefont{Dalvit}},
  \bibnamefont{and} \bibinfo{author}{\bibfnamefont{P.~W.}
  \bibnamefont{Milonni}}, \bibinfo{journal}{Phys. Rev.~A}
  \textbf{\bibinfo{volume}{78}}, \bibinfo{pages}{032117}
  (\bibinfo{year}{2008}{\natexlab{b}}).

\bibitem[{\citenamefont{Dzyaloshinski{\u{\i}}
  et~al.}(1961)\citenamefont{Dzyaloshinski{\u{\i}}, Lifshitz, and
  Pitaevski{\u{\i}}}}]{Dzyaloshinskii61}
\bibinfo{author}{\bibfnamefont{I.~E.} \bibnamefont{Dzyaloshinski{\u{\i}}}},
  \bibinfo{author}{\bibfnamefont{E.~M.} \bibnamefont{Lifshitz}},
  \bibnamefont{and} \bibinfo{author}{\bibfnamefont{L.~P.}
  \bibnamefont{Pitaevski{\u{\i}}}}, \bibinfo{journal}{Adv. Phys.}
  \textbf{\bibinfo{volume}{10}}, \bibinfo{pages}{165} (\bibinfo{year}{1961}).

\bibitem[{\citenamefont{Zhao et~al.}(2009)\citenamefont{Zhao, Zhou, Koschny,
  Economou, and Soukoulis}}]{Zhao09}
\bibinfo{author}{\bibfnamefont{R.}~\bibnamefont{Zhao}},
  \bibinfo{author}{\bibfnamefont{J.}~\bibnamefont{Zhou}},
  \bibinfo{author}{\bibfnamefont{T.}~\bibnamefont{Koschny}},
  \bibinfo{author}{\bibfnamefont{E.~N.} \bibnamefont{Economou}},
  \bibnamefont{and} \bibinfo{author}{\bibfnamefont{C.~M.}
  \bibnamefont{Soukoulis}}, \bibinfo{journal}{Phys. Rev. Lett.}
  \textbf{\bibinfo{volume}{103}}, \bibinfo{pages}{103602}
  (\bibinfo{year}{2009}).

\bibitem[{\citenamefont{Rahi et~al.}(2010)\citenamefont{Rahi, Kardar, and
  Emig}}]{Rahi10:PRL}
\bibinfo{author}{\bibfnamefont{S.~J.} \bibnamefont{Rahi}},
  \bibinfo{author}{\bibfnamefont{M.}~\bibnamefont{Kardar}}, \bibnamefont{and}
  \bibinfo{author}{\bibfnamefont{T.}~\bibnamefont{Emig}},
  \bibinfo{journal}{Phys. Rev. Lett.} \textbf{\bibinfo{volume}{105}},
  \bibinfo{pages}{070404} (\bibinfo{year}{2010}).

\bibitem[{\citenamefont{McCauley
  et~al.}(2010{\natexlab{a}})\citenamefont{McCauley, Zhao, Reid, Rodriguez,
  Zhuo, Rosa, Joannopoulos, Dalvit, Soukoulis, and Johnson}}]{McCauleyZh10}
\bibinfo{author}{\bibfnamefont{A.~P.} \bibnamefont{McCauley}},
  \bibinfo{author}{\bibfnamefont{R.}~\bibnamefont{Zhao}},
  \bibinfo{author}{\bibfnamefont{M.~T.~H.} \bibnamefont{Reid}},
  \bibinfo{author}{\bibfnamefont{A.~W.} \bibnamefont{Rodriguez}},
  \bibinfo{author}{\bibfnamefont{J.}~\bibnamefont{Zhuo}},
  \bibinfo{author}{\bibfnamefont{F.~S.~S.} \bibnamefont{Rosa}},
  \bibinfo{author}{\bibfnamefont{J.~D.} \bibnamefont{Joannopoulos}},
  \bibinfo{author}{\bibfnamefont{D.~A.~R.} \bibnamefont{Dalvit}},
  \bibinfo{author}{\bibfnamefont{C.~M.} \bibnamefont{Soukoulis}},
  \bibnamefont{and} \bibinfo{author}{\bibfnamefont{S.~G.}
  \bibnamefont{Johnson}}, \bibinfo{journal}{Phys. Rev.~B}
  \textbf{\bibinfo{volume}{82}}, \bibinfo{pages}{165108}
  (\bibinfo{year}{2010}{\natexlab{a}}).

\bibitem[{\citenamefont{van Enk}(1995)}]{Enk95:torque}
\bibinfo{author}{\bibfnamefont{S.~J.} \bibnamefont{van Enk}},
  \bibinfo{journal}{Phys. Rev.~A} \textbf{\bibinfo{volume}{52}},
  \bibinfo{pages}{2569} (\bibinfo{year}{1995}).

\bibitem[{\citenamefont{Shao et~al.}(2005)\citenamefont{Shao, Tong, and
  Luo}}]{Shao05}
\bibinfo{author}{\bibfnamefont{C.-G.} \bibnamefont{Shao}},
  \bibinfo{author}{\bibfnamefont{A.-H.} \bibnamefont{Tong}}, \bibnamefont{and}
  \bibinfo{author}{\bibfnamefont{J.}~\bibnamefont{Luo}},
  \bibinfo{journal}{Phys. Rev.~A} \textbf{\bibinfo{volume}{72}},
  \bibinfo{pages}{022102} (\bibinfo{year}{2005}).

\bibitem[{\citenamefont{Munday et~al.}(2005)\citenamefont{Munday, Iannuzzi,
  Barash, and Capasso}}]{Munday05}
\bibinfo{author}{\bibfnamefont{J.~N.} \bibnamefont{Munday}},
  \bibinfo{author}{\bibfnamefont{D.}~\bibnamefont{Iannuzzi}},
  \bibinfo{author}{\bibfnamefont{Y.}~\bibnamefont{Barash}}, \bibnamefont{and}
  \bibinfo{author}{\bibfnamefont{F.}~\bibnamefont{Capasso}},
  \bibinfo{journal}{Phys. Rev.~A} \textbf{\bibinfo{volume}{71}},
  \bibinfo{pages}{042102} (\bibinfo{year}{2005}).

\bibitem[{\citenamefont{Kenneth and Nussinov}(2000)}]{Kenneth00:torque}
\bibinfo{author}{\bibfnamefont{O.}~\bibnamefont{Kenneth}} \bibnamefont{and}
  \bibinfo{author}{\bibfnamefont{S.}~\bibnamefont{Nussinov}},
  \bibinfo{journal}{arXiv:hep-th/0001045}  (\bibinfo{year}{2000}).

\bibitem[{\citenamefont{Rodriguez et~al.}(2011)\citenamefont{Rodriguez, Woolf,
  Hui, Iwase, McCauley, Capasso, Loncar, and Johnson}}]{Rodriguez11:OL}
\bibinfo{author}{\bibfnamefont{A.~W.} \bibnamefont{Rodriguez}},
  \bibinfo{author}{\bibfnamefont{D.}~\bibnamefont{Woolf}},
  \bibinfo{author}{\bibfnamefont{P.-C.} \bibnamefont{Hui}},
  \bibinfo{author}{\bibfnamefont{E.}~\bibnamefont{Iwase}},
  \bibinfo{author}{\bibfnamefont{A.~P.} \bibnamefont{McCauley}},
  \bibinfo{author}{\bibfnamefont{F.}~\bibnamefont{Capasso}},
  \bibinfo{author}{\bibfnamefont{M.}~\bibnamefont{Loncar}}, \bibnamefont{and}
  \bibinfo{author}{\bibfnamefont{S.~G.} \bibnamefont{Johnson}},
  \bibinfo{journal}{arXiv} \textbf{\bibinfo{volume}{1101.4237}}
  (\bibinfo{year}{2011}).

\bibitem[{\citenamefont{Morecroft et~al.}(2009)\citenamefont{Morecroft, Yang,
  Schuster, Berggren, Xia, Wu, and Williams}}]{Morecroft2009}
\bibinfo{author}{\bibfnamefont{D.}~\bibnamefont{Morecroft}},
  \bibinfo{author}{\bibfnamefont{J.~K.~W.} \bibnamefont{Yang}},
  \bibinfo{author}{\bibfnamefont{S.}~\bibnamefont{Schuster}},
  \bibinfo{author}{\bibfnamefont{K.~K.} \bibnamefont{Berggren}},
  \bibinfo{author}{\bibfnamefont{Q.}~\bibnamefont{Xia}},
  \bibinfo{author}{\bibfnamefont{W.}~\bibnamefont{Wu}}, \bibnamefont{and}
  \bibinfo{author}{\bibfnamefont{R.~S.} \bibnamefont{Williams}},
  \bibinfo{journal}{J. Vac. Sci. Technol. B} \textbf{\bibinfo{volume}{27}},
  \bibinfo{pages}{2837} (\bibinfo{year}{2009}).

\bibitem[{\citenamefont{Lambrecht and Marachevsky}(2008)}]{Lambrecht09}
\bibinfo{author}{\bibfnamefont{A.}~\bibnamefont{Lambrecht}} \bibnamefont{and}
  \bibinfo{author}{\bibfnamefont{V.~N.} \bibnamefont{Marachevsky}},
  \bibinfo{journal}{Phys. Rev. Lett.} \textbf{\bibinfo{volume}{101}},
  \bibinfo{pages}{160403} (\bibinfo{year}{2008}).

\bibitem[{\citenamefont{Davids et~al.}(2010)\citenamefont{Davids, Intravaia,
  Rosa, and Dalvit}}]{Davids10}
\bibinfo{author}{\bibfnamefont{P.~S.} \bibnamefont{Davids}},
  \bibinfo{author}{\bibfnamefont{F.}~\bibnamefont{Intravaia}},
  \bibinfo{author}{\bibfnamefont{F.~S.~S.} \bibnamefont{Rosa}},
  \bibnamefont{and} \bibinfo{author}{\bibfnamefont{D.~A.~R.}
  \bibnamefont{Dalvit}}, \bibinfo{journal}{Phys. Rev.~A}
  \textbf{\bibinfo{volume}{82}}, \bibinfo{pages}{062111}
  (\bibinfo{year}{2010}).

\bibitem[{\citenamefont{Reid et~al.}(2009)\citenamefont{Reid, Rodriguez, White,
  and Johnson}}]{ReidRo09}
\bibinfo{author}{\bibfnamefont{M.~T.~H.} \bibnamefont{Reid}},
  \bibinfo{author}{\bibfnamefont{A.~W.} \bibnamefont{Rodriguez}},
  \bibinfo{author}{\bibfnamefont{J.}~\bibnamefont{White}}, \bibnamefont{and}
  \bibinfo{author}{\bibfnamefont{S.~G.} \bibnamefont{Johnson}},
  \bibinfo{journal}{Phys. Rev. Lett.} \textbf{\bibinfo{volume}{103}},
  \bibinfo{pages}{040401} (\bibinfo{year}{2009}).

\bibitem[{\citenamefont{Reid~et. al.}(2011)}]{Reid11:2D}
\bibinfo{author}{\bibfnamefont{M.~T.~H.} \bibnamefont{Reid~et. al.}},
  \bibinfo{journal}{In Preparation}  (\bibinfo{year}{2011}).

\bibitem[{\citenamefont{Rodriguez et~al.}(2009)\citenamefont{Rodriguez,
  McCauley, Joannopoulos, and Johnson}}]{RodriguezMc09:PRA}
\bibinfo{author}{\bibfnamefont{A.~W.} \bibnamefont{Rodriguez}},
  \bibinfo{author}{\bibfnamefont{A.~P.} \bibnamefont{McCauley}},
  \bibinfo{author}{\bibfnamefont{J.~D.} \bibnamefont{Joannopoulos}},
  \bibnamefont{and} \bibinfo{author}{\bibfnamefont{S.~G.}
  \bibnamefont{Johnson}}, \bibinfo{journal}{Phys. Rev.~A}
  \textbf{\bibinfo{volume}{80}}, \bibinfo{pages}{012115}
  (\bibinfo{year}{2009}).

\bibitem[{\citenamefont{McCauley
  et~al.}(2010{\natexlab{b}})\citenamefont{McCauley, Rodriguez, Joannopoulos,
  and Johnson}}]{McCauleyRo10:PRA}
\bibinfo{author}{\bibfnamefont{A.~P.} \bibnamefont{McCauley}},
  \bibinfo{author}{\bibfnamefont{A.~W.} \bibnamefont{Rodriguez}},
  \bibinfo{author}{\bibfnamefont{J.~D.} \bibnamefont{Joannopoulos}},
  \bibnamefont{and} \bibinfo{author}{\bibfnamefont{S.~G.}
  \bibnamefont{Johnson}}, \bibinfo{journal}{Phys. Rev.~A}
  \textbf{\bibinfo{volume}{81}}, \bibinfo{pages}{012119}
  (\bibinfo{year}{2010}{\natexlab{b}}).

\bibitem[{\citenamefont{Emig et~al.}(2007)\citenamefont{Emig, Graham, Jaffe,
  and Kardar}}]{Emig07}
\bibinfo{author}{\bibfnamefont{T.}~\bibnamefont{Emig}},
  \bibinfo{author}{\bibfnamefont{N.}~\bibnamefont{Graham}},
  \bibinfo{author}{\bibfnamefont{R.~L.} \bibnamefont{Jaffe}}, \bibnamefont{and}
  \bibinfo{author}{\bibfnamefont{M.}~\bibnamefont{Kardar}},
  \bibinfo{journal}{Phys. Rev. Lett.} \textbf{\bibinfo{volume}{99}},
  \bibinfo{pages}{170403} (\bibinfo{year}{2007}).

\bibitem[{\citenamefont{Kenneth and Klich}(2008)}]{Kenneth08}
\bibinfo{author}{\bibfnamefont{O.}~\bibnamefont{Kenneth}} \bibnamefont{and}
  \bibinfo{author}{\bibfnamefont{I.}~\bibnamefont{Klich}},
  \bibinfo{journal}{Phys. Rev.~B} \textbf{\bibinfo{volume}{78}},
  \bibinfo{pages}{014103} (\bibinfo{year}{2008}).

\bibitem[{\citenamefont{Maia~Neto et~al.}(2008)\citenamefont{Maia~Neto,
  Lambrecht, and Reynaud}}]{Neto08}
\bibinfo{author}{\bibfnamefont{P.~A.} \bibnamefont{Maia~Neto}},
  \bibinfo{author}{\bibfnamefont{A.}~\bibnamefont{Lambrecht}},
  \bibnamefont{and} \bibinfo{author}{\bibfnamefont{S.}~\bibnamefont{Reynaud}},
  \bibinfo{journal}{Phys. Rev.~A} \textbf{\bibinfo{volume}{78}},
  \bibinfo{pages}{012115} (\bibinfo{year}{2008}).

\bibitem[{\citenamefont{Parsegian and Weiss}(1972)}]{Parsegian72}
\bibinfo{author}{\bibfnamefont{V.~A.} \bibnamefont{Parsegian}}
  \bibnamefont{and} \bibinfo{author}{\bibfnamefont{G.~H.} \bibnamefont{Weiss}},
  \bibinfo{journal}{J. Adh.} \textbf{\bibinfo{volume}{3}}, \bibinfo{pages}{259}
  (\bibinfo{year}{1972}).

\bibitem{Barash75} Y. Barash and V.L. Ginzburg, Usp. Fiz. Nauk {\bf 116}, 5 (1975) 
[Sov. Phys. Usp. {\bf 18}, 305 (1975)].


\bibitem{Barash78} Y. Barash, Izv. Vyssh. Uchebn. Zaved. Radiofiz. {\bf 21}, 1637 (1978)
[Radiophysics and Quantum Electronics {\bf 21}, 1138 (1978)].  


\bibitem[{\citenamefont{Philbin and Leonhardt}(2008)}]{Philbin08}
\bibinfo{author}{\bibfnamefont{T.~G.} \bibnamefont{Philbin}} \bibnamefont{and}
  \bibinfo{author}{\bibfnamefont{U.}~\bibnamefont{Leonhardt}},
  \bibinfo{journal}{Phys. Rev.~A} \textbf{\bibinfo{volume}{78}},
  \bibinfo{pages}{042107} (\bibinfo{year}{2008}).

\bibitem[{\citenamefont{Milling et~al.}(1996)\citenamefont{Milling, Mulvaney,
  and Larson}}]{Milling96}
\bibinfo{author}{\bibfnamefont{A.}~\bibnamefont{Milling}},
  \bibinfo{author}{\bibfnamefont{P.}~\bibnamefont{Mulvaney}}, \bibnamefont{and}
  \bibinfo{author}{\bibfnamefont{I.}~\bibnamefont{Larson}},
  \bibinfo{journal}{J. Colloid Interface Sci.} \textbf{\bibinfo{volume}{180}},
  \bibinfo{pages}{460} (\bibinfo{year}{1996}).

\bibitem[{\citenamefont{Bergstrom}(1997)}]{Bergstrom97}
\bibinfo{author}{\bibfnamefont{L.}~\bibnamefont{Bergstrom}},
  \bibinfo{journal}{Adv. Colloid and Interface Science}
  \textbf{\bibinfo{volume}{70}}, \bibinfo{pages}{125} (\bibinfo{year}{1997}).

\bibitem[{\citenamefont{Kenneth and Klich}(2006)}]{KennethKl06}
\bibinfo{author}{\bibfnamefont{O.}~\bibnamefont{Kenneth}} \bibnamefont{and}
  \bibinfo{author}{\bibfnamefont{I.}~\bibnamefont{Klich}},
  \bibinfo{journal}{Phys. Rev. Lett.} \textbf{\bibinfo{volume}{97}},
  \bibinfo{pages}{160401} (\bibinfo{year}{2006}).

\bibitem[{\citenamefont{Pendry et~al.}(1996)\citenamefont{Pendry, Holden,
  Stewart, and Youngs}}]{Pendry96}
\bibinfo{author}{\bibfnamefont{J.~B.} \bibnamefont{Pendry}},
  \bibinfo{author}{\bibfnamefont{A.~J.} \bibnamefont{Holden}},
  \bibinfo{author}{\bibfnamefont{W.~J.} \bibnamefont{Stewart}},
  \bibnamefont{and} \bibinfo{author}{\bibfnamefont{I.}~\bibnamefont{Youngs}},
  \bibinfo{journal}{Phys. Rev. Lett.} \textbf{\bibinfo{volume}{76}},
  \bibinfo{pages}{4773} (\bibinfo{year}{1996}).

\bibitem[{\citenamefont{Maghrebi}(2010)}]{Maghrebi10:arxiv}
\bibinfo{author}{\bibfnamefont{M.~F.} \bibnamefont{Maghrebi}},
  \bibinfo{journal}{arXiv} \textbf{\bibinfo{volume}{1012.1060}}
  (\bibinfo{year}{2010}).

\bibitem[{\citenamefont{Lamoreaux}(1997)}]{Lamoreaux97}
\bibinfo{author}{\bibfnamefont{S.~K.} \bibnamefont{Lamoreaux}},
  \bibinfo{journal}{Phys. Rev. Lett.} \textbf{\bibinfo{volume}{78}},
  \bibinfo{pages}{5} (\bibinfo{year}{1997}).

\bibitem[{\citenamefont{Mohideen and Roy}(1998)}]{moh1}
\bibinfo{author}{\bibfnamefont{U.}~\bibnamefont{Mohideen}} \bibnamefont{and}
  \bibinfo{author}{\bibfnamefont{A.}~\bibnamefont{Roy}},
  \bibinfo{journal}{Phys. Rev. Lett.} \textbf{\bibinfo{volume}{81}},
  \bibinfo{pages}{4549} (\bibinfo{year}{1998}).

\bibitem[{\citenamefont{Brown-Hayes et~al.}(2005)\citenamefont{Brown-Hayes,
  Dalvit, Mazzitelli, Kim, and Onofrio}}]{Brown-Hayes05}
\bibinfo{author}{\bibfnamefont{M.}~\bibnamefont{Brown-Hayes}},
  \bibinfo{author}{\bibfnamefont{D.~A.~R.} \bibnamefont{Dalvit}},
  \bibinfo{author}{\bibfnamefont{F.~D.} \bibnamefont{Mazzitelli}},
  \bibinfo{author}{\bibfnamefont{W.~J.} \bibnamefont{Kim}}, \bibnamefont{and}
  \bibinfo{author}{\bibfnamefont{R.}~\bibnamefont{Onofrio}},
  \bibinfo{journal}{Phys. Rev.~A} \textbf{\bibinfo{volume}{72}},
  \bibinfo{pages}{052102} (\bibinfo{year}{2005}).

\bibitem[{\citenamefont{Wei et~al.}(2010)\citenamefont{Wei, Dalvit, Lombardo,
  Mazzitelli, and Onofrio}}]{Wei10}
\bibinfo{author}{\bibfnamefont{Q.}~\bibnamefont{Wei}},
  \bibinfo{author}{\bibfnamefont{D.~A.~R.} \bibnamefont{Dalvit}},
  \bibinfo{author}{\bibfnamefont{F.~C.} \bibnamefont{Lombardo}},
  \bibinfo{author}{\bibfnamefont{F.~D.} \bibnamefont{Mazzitelli}},
  \bibnamefont{and} \bibinfo{author}{\bibfnamefont{R.}~\bibnamefont{Onofrio}},
  \bibinfo{journal}{Phys. Rev.~A} \textbf{\bibinfo{volume}{81}},
  \bibinfo{pages}{052115} (\bibinfo{year}{2010}).

\bibitem[{\citenamefont{de~Man et~al.}(2010{\natexlab{a}})\citenamefont{de~Man,
  Heeck, Wijngaarden, and Iannuzzi}}]{Man10}
\bibinfo{author}{\bibfnamefont{S.}~\bibnamefont{de~Man}},
  \bibinfo{author}{\bibfnamefont{K.}~\bibnamefont{Heeck}},
  \bibinfo{author}{\bibfnamefont{R.~J.} \bibnamefont{Wijngaarden}},
  \bibnamefont{and} \bibinfo{author}{\bibfnamefont{D.}~\bibnamefont{Iannuzzi}},
  \bibinfo{journal}{J. Vac. Sci. Technol. B} \textbf{\bibinfo{volume}{28}}
  (\bibinfo{year}{2010}{\natexlab{a}}).

\bibitem[{\citenamefont{de~Man et~al.}(2010{\natexlab{b}})\citenamefont{de~Man,
  Heeck, and Iannuzzi}}]{Man10:PRA}
\bibinfo{author}{\bibfnamefont{S.}~\bibnamefont{de~Man}},
  \bibinfo{author}{\bibfnamefont{K.}~\bibnamefont{Heeck}}, \bibnamefont{and}
  \bibinfo{author}{\bibfnamefont{D.}~\bibnamefont{Iannuzzi}},
  \bibinfo{journal}{Phys. Rev.~A} \textbf{\bibinfo{volume}{82}},
  \bibinfo{pages}{062512} (\bibinfo{year}{2010}{\natexlab{b}}).

\bibitem[{\citenamefont{Johnson}(2010)}]{Steven11:review}
\bibinfo{author}{\bibfnamefont{S.~G.} \bibnamefont{Johnson}},
  \bibinfo{journal}{arXiv:quant-ph/1007.0966}  (\bibinfo{year}{2010}),
  \bibinfo{note}{to appear in "Lecture Notes in Physics: {Casimir} physics
  (Springer, eds. D.~A.~R. Dalvit, P.~W. Milonni, D.~C. Roberts, and F.~S.~S.
  Rosa}.

\bibitem[{\citenamefont{Rodriguez
  et~al.}(2007{\natexlab{b}})\citenamefont{Rodriguez, Ibanescu, Iannuzzi,
  Joannopoulos, and Johnson}}]{Rodriguez07:PRA}
\bibinfo{author}{\bibfnamefont{A.}~\bibnamefont{Rodriguez}},
  \bibinfo{author}{\bibfnamefont{M.}~\bibnamefont{Ibanescu}},
  \bibinfo{author}{\bibfnamefont{D.}~\bibnamefont{Iannuzzi}},
  \bibinfo{author}{\bibfnamefont{J.~D.} \bibnamefont{Joannopoulos}},
  \bibnamefont{and} \bibinfo{author}{\bibfnamefont{S.~G.}
  \bibnamefont{Johnson}}, \bibinfo{journal}{Phys. Rev.~A}
  \textbf{\bibinfo{volume}{76}}, \bibinfo{pages}{032106}
  (\bibinfo{year}{2007}{\natexlab{b}}).

\bibitem[{\citenamefont{Pasquali and Maggs}(2009)}]{Pasquali09}
\bibinfo{author}{\bibfnamefont{S.}~\bibnamefont{Pasquali}} \bibnamefont{and}
  \bibinfo{author}{\bibfnamefont{A.~C.} \bibnamefont{Maggs}},
  \bibinfo{journal}{Phys. Rev. A.} \textbf{\bibinfo{volume}{79}},
  \bibinfo{pages}{020102({R})} (\bibinfo{year}{2009}).

\bibitem[{\citenamefont{Moharam et~al.}(1995)\citenamefont{Moharam, Grann,
  Pommet, and Gaylord}}]{Moharam95}
\bibinfo{author}{\bibfnamefont{M.~G.} \bibnamefont{Moharam}},
  \bibinfo{author}{\bibfnamefont{E.~B.} \bibnamefont{Grann}},
  \bibinfo{author}{\bibfnamefont{D.~A.} \bibnamefont{Pommet}},
  \bibnamefont{and} \bibinfo{author}{\bibfnamefont{T.~K.}
  \bibnamefont{Gaylord}}, \bibinfo{journal}{J.~Opt. Soc. Am.~A}
  \textbf{\bibinfo{volume}{12}}, \bibinfo{pages}{1077} (\bibinfo{year}{1995}).

\bibitem[{\citenamefont{Bienstman and Baets}(2001)}]{Bienstman01}
\bibinfo{author}{\bibfnamefont{P.}~\bibnamefont{Bienstman}} \bibnamefont{and}
  \bibinfo{author}{\bibfnamefont{R.}~\bibnamefont{Baets}},
  \bibinfo{journal}{Optical and Quantum Electron.}
  \textbf{\bibinfo{volume}{33}}, \bibinfo{pages}{327} (\bibinfo{year}{2001}).

\bibitem[{\citenamefont{Stratton}(1941)}]{Stratton41}
\bibinfo{author}{\bibfnamefont{J.~A.} \bibnamefont{Stratton}},
  \emph{\bibinfo{title}{Electromagnetic Theory}}
  (\bibinfo{publisher}{McGraw-Hill}, \bibinfo{address}{New York},
  \bibinfo{year}{1941}).

\bibitem[{\citenamefont{B{\"{u}}scher and Emig}(2004)}]{emig04_2}
\bibinfo{author}{\bibfnamefont{R.}~\bibnamefont{B{\"{u}}scher}}
  \bibnamefont{and} \bibinfo{author}{\bibfnamefont{T.}~\bibnamefont{Emig}},
  \bibinfo{journal}{Phys. Rev.~A} \textbf{\bibinfo{volume}{69}},
  \bibinfo{pages}{062101} (\bibinfo{year}{2004}).

\bibitem[{\citenamefont{Landau et~al.}(1960)\citenamefont{Landau, Lifshitz, and
  Pitaevski{\u{\i}}}}]{Landau:EM}
\bibinfo{author}{\bibfnamefont{L.~D.} \bibnamefont{Landau}},
  \bibinfo{author}{\bibfnamefont{E.~M.} \bibnamefont{Lifshitz}},
  \bibnamefont{and} \bibinfo{author}{\bibfnamefont{L.~P.}
  \bibnamefont{Pitaevski{\u{\i}}}}, \emph{\bibinfo{title}{Electrodynamics of
  Continuous Media}}, vol.~\bibinfo{volume}{8} (\bibinfo{publisher}{Pergamon
  Press}, \bibinfo{address}{Oxford}, \bibinfo{year}{1960}).

\end{thebibliography}

\end{document}